\documentclass{aastex6}

\usepackage{color} 
\usepackage{bm}

\fullcollaborationName{}

\begin{document}


\title{Ring Formation around Giant Planets by Tidal Disruption of a Single Passing Large Kuiper Belt Object}


\author{Ryuki Hyodo\altaffilmark{1,2}, S{\'e}bastien Charnoz\altaffilmark{1,3}, Keiji Ohtsuki\altaffilmark{2} \& Hidenori Genda\altaffilmark{4}}


\altaffiltext{1}{Institut de Physique du Globe, Paris 75005, France}
\altaffiltext{2}{Department of Planetology, Kobe University, Kobe 657-8501, Japan}
\altaffiltext{3}{Université Paris Diderot, Paris 75013, France}
\altaffiltext{4}{Earth-Life Science Institute, Tokyo Institute of Technology, Tokyo 152-8550, Japan}

\begin{abstract}
The origin of rings around giant planets remains elusive. Saturn's rings are massive and made of 90-95\% of water ice with a mass of $\sim 10^{19}$ kg. In contrast, the much less massive rings of Uranus and Neptune are dark and likely to have higher rock fraction. According to the so-called "Nice model", at the time of the Late Heavy Bombardment, giant planets could have experienced a significant number of close encounters with bodies scattered from the primordial Kuiper Belt. This belt could have been massive in the past and may have contained a larger number of big objects ($M_{\rm body}=10^{22}$kg) than what is currently observed in the Kuiper Belt. Here we investigate, for the first time, the tidal disruption of a passing object, including the subsequent formation of planetary rings. First, we perform SPH simulations of the tidal destruction of big differentiated objects ($M_{\rm body}=10^{21}$ and $10^{23}$kg) that experience close encounters with Saturn or Uranus. We find that about $0.1-10$\% of the mass of the passing body is gravitationally captured around the planet. However, these fragments are initially big chunks and have highly eccentric orbits around the planet. In order to see their long-term evolution, we perform N-body simulations including the planet's oblateness up to $J_4$ starting with data obtained from the SPH simulations. Our N-body simulations show that the chunks are tidally destroyed during their next several orbits and become collections of smaller particles. Their individual orbits then start to precess incoherently around the planet's equator, which enhances their encounter velocities on longer-term evolution, resulting in more destructive impacts. These collisions would damp their eccentricities resulting in a progressive collapse of the debris cloud into a thin equatorial and low-eccentricity ring. These high energy impacts are expected to be catastrophic enough to produce small particles. Our numerical results also show that the mass of formed rings is large enough to explain current rings including inner regular satellites around Saturn and Uranus. In the case of Uranus, a body can go deeper inside the planet's Roche limit resulting in a more efficient capture of rocky material compared to Saturn's case in which mostly ice is captured. Thus, our results can naturally explain the compositional difference between the rings of Saturn, Uranus and Neptune. 

\end{abstract}




\section{Introduction} \label{sec:intro}
The origin of planetary rings is still a debated question. Saturn's main rings are unique as they are made of 90-95\% water ice \citep{Cuz98, Pou03, Nic05} with a mass of $\sim 10^{19}$ kg \citep{Esp83, Cha09}. In contrast, the much less massive rings of Uranus and Neptune are dark and likely to have a higher rock content \citep{Tis13} than Saturn's rings. Whereas dusty Saturn's E and G rings are likely to be formed via the destruction or surface erosion of the nearby present satellites \citep{Esp93, Col94, Bur01, Hed07, Por06}, Saturn's main rings cannot result from the same process as there is no obvious source of material to feed them today. Note, however, that a recent study shows that the origin of Saturn's F ring and Uranian $\epsilon$ ring could be a natural consequence of the collisional destruction between small satellites just outside the main rings \citep{Hyo15b} that is formed by the spreading of ancient rings \citep{Cha10, Cri12, Hyo15a}.\\

Several ring formation scenarios have been proposed for massive rings, like those of Saturn: (1) primordial satellite collisional destruction by passing comet \citep{Pol73, Pol75, Har84}, (2) tidal destruction of a primordial satellite at the Roche Limit after inward migration due to tidal interaction with the circum-Saturn gas disk \citep{Can10} and (3) tidal disruption of passing objects \citep{Don91}. The inward migration of a primordial Titan-sized satellite and the removal of only its pure icy mantle could beautifully explain the silicate deficit of Saturn's rings. However, it requires some fine-tuning of the timing of the event (likely at the end of the evolution of the circum-Saturn circumplanetary disk) so that the disk is still massive enough to allow inward migration of the satellite, but light enough in order to prevent the rapid infall of debris into the planet because of gas drag. In addition, it would be difficult to directly form centimetre- to meter-sized particles that are currently seen in Saturn's main rings by tidal destruction alone. \cite{Can10} proposes collision between fragments can form small particles, but detailed studies are still required. Tidal disruption of a passing differentiated object could also potentially explain the high ice/rock fraction by capturing only the icy mantle of the incoming body and letting the remnant core escape from Saturn's gravity field. However, so far this scenario has scarcely been studied and only used a simplified analytical model of a homogeneous body \citep{Don91, Cha09}. Thus, direct numerical simulation of the tidal splitting of a big differentiated body is necessary now to investigate this scenario in detail. In addition, even though some mass capturing occurs, the fragments are expected to have highly eccentric orbits around the planet and the long-term evolution of such fragments remains unclear, in particular by which process a ring of cm-sized particles forms. In this work, we investigate, for the first time,  the details of tidal disruption of a passing large differentiated object and the long-term fate of its debris by using direct simulations.\\

Such an event may have occurred, with the most probability, either during the phase of planet formation where the giant-planets' cores are expected to scatter the neighbouring planetesimals efficiently, or later during the Late Heavy Bombardment (LHB). The well known "Nice model" explains not only the Lunar cataclysm or today's orbital architecture of giant planets \citep{Gom05, Tsi05}, but also the implantation of Jupiter's Trojan asteroids as well as the irregular satellites around giant planets \citep{Mor05, Nes07}. During this instability phase, giant planets could have experienced a significant number of close encounters with bodies scattered from the primordial Kuiper Belt that surrounded the giant planets. This belt could have been significantly massive and may have contained a larger number of big objects than what is currently observed in the Kuiper Belt \citep{Lev08,Nes16}. \cite{Cha09} estimated an encounter rate of the primordial Kuiper Belt Objects (KBOs) with the giant planets during the LHB and investigated the captured mass around giant planets using a simplified analytical model assuming homogeneous small bodies like comets. The flux of such undifferentiated small bodies is enormous and isotropic, and thus the average angular momentum of captured mass should be almost zero, resulting in no contribution to the formation of the rings. However, only a single tidal disruption of a large object that deviates from the average could decide the story, and such large objects could be expected to be differentiated like Pluto.\\

Here we use two different direct simulations and investigate successive processes from tidal destruction of a passing object to the possible formation of planetary rings (Figure \ref{summary}). Our work will address (1) the captured mass as well as its ice/silicate fraction due to the tidal disruption of differentiated bodies at different planets, (2) the orbits of the captured fragments, and (3) the long-term orbital and collisional evolution of the captured fragments. To do so, we first investigate the physics of tidal disruption of a differentiated object that is initially on a hyperbolic orbit about Saturn or Uranus, and calculate how much mass is gravitationally captured around these planets by using smooth particle hydrodynamics (SPH) simulations. Then we perform direct N-body simulations in order to see the longer-term evolution of such captured fragments around Saturn, including the effect of oblateness potential of the planet. In section 2, in light of our newly derived semi-analytical models that take into account spin and self-gravity of a differentiated object, we briefly review a previous analytical formula of the capture efficiency of tidal disruption. In section 3, we explain our SPH method and model. In section 4, we show the results from SPH simulations and discuss the mass capture efficiency as well as orbits of captured fragments. In section 5, using the data obtained from SPH simulations as initial conditions, we perform N-body simulations of subsequent long-term evolution of captured fragments. Then, using results of the simulations as well as analytic estimation, we discuss the fate of the captured fragments. Section 6 summarises our results and discusses the origin of planetary rings.

\section{Physical Argument}

\subsection{Previous Analytical Model}
\cite{Don91} derived the mass that is captured on bound orbits around a planet during tidal disruption at close encounters. Assuming uniform energy distribution across the body, the amplitude of energy variation is, 
\begin{equation}
	\Delta E = \frac{GM_0}{q} \times \frac{R}{q}
\end{equation}
where $G$, $M_0$, $q$ and $R$ are the gravitational constant, the mass of the planet, pericenter distance and radius of the body, respectively. The mass fraction that is captured with $E<E_{\rm sta}$ is obtained assuming energy distribution is uniform between $-0.9\Delta E + 0.5 v^2_{\rm inf}$ and $0.9\Delta E + 0.5 v^2_{\rm inf}$ as

\begin{equation}
	\eta = \frac{M_{\rm bound}}{M_{\rm object}}=\frac{0.9\Delta E + E_{\rm sta} - 0.5v^2_{\rm inf}}{1.8\Delta E}
\label{Dones1991}
\end{equation}
where $v_{\rm inf}$ is the velocity of the object at infinity and we take $E_{\rm sta}=-GM_0/R_{\rm Hill}$ where $R_{\rm Hill}$ is the planet's Hill radius. Equation \ref{Dones1991} (hereafter we call this Dones' formula) has been used to estimate mass captured around giant planets in \cite{Cha09}. However, it has not been well established if this formula is applicable in any case.

\subsection{Semi-analytical Model: Effects of Spin and Self-gravity}
The above Dones' formula considers only ballistic orbits for all constituent particles of the body. However, the spin state of the body as well as the body's self-gravity could play an important role in the capture efficiency, especially when the body is massive. Here, we consider a spherical body located at pericenter during its close encounter with Saturn (mass $M_{\rm S}$). We three-dimensionally divide a cubic box that contains the body into $N^3$ cells by using a cartesian grid, with $N=100$ along 1-dimension. The box width is $1.1$ times the diameter of the spherical body and the object's center of mass is located at the center of the box. Each cell $i$ within the body has its mass $m_{i}$, position $r_{i}$ and velocity $v_{i}$ relative to Saturn's center.  Then, we calculate the energy $E_{i}=\frac{1}{2}m_{i}v_{i}^2-\frac{Gm_{i}M_{\rm S}}{r_{i}}$ of each cell, and assume any cell that satisfies $E_{i}<0$ is gravitationally captured around Saturn. The body is a differentiated spherical body of mass $M_{\rm body}=10^{21}$kg or $10^{23}$kg and the mass fraction of the core is 0.5. The densities of the core and the mantle are $3000$kg/m$^{3}$ and $900$kg/m$^{3}$, respectively. Using the above procedure, we calculate the capture efficiency at different pericenter distances ranging from $25 \times 10^{3}$km to $100 \times 10^{3}$km . Figure \ref{modeled_body} shows an example of our modelled passing body with pericenter distance $70 \times 10^{3}$km and velocity at infinity $v_{\rm inf}=3$km/s.

\subsubsection{Effects of Spin}
Figure \ref{effect_of_spin} shows the capture efficiency obtained by our semi-analytical model described above. We include the spin velocity around the object's center of mass with spin period $T_{\rm spin}=8$h in the prograde or retrograde direction.  Here, prograde rotation means that the body rotates in the same direction as its orbit. Closer to Saturn, the energy decreases due to Saturn's potential field. Therefore, inside the body, cells closer to Saturn are more prone to having negative energy. However, with retrograde spin, the radial inner half of the body has a higher spin velocity in the  same direction as its orbital velocity at pericenter, and thus achieves larger relative velocity at pericenter compared to the no spin case. Therefore, capture efficiency becomes lower in the case of retrograde spin. In contrast, when a body has prograde spin, the radial inner half of the body has a smaller relative velocity with respect to Saturn, resulting in a larger capture efficiency as seen in Figure \ref{effect_of_spin}.

\subsubsection{Effects of Self-gravity}
During the tidal disruption of a passing body on a parabolic orbit, Dones' formula assumes that the body is instantaneously destroyed and that the inner half is captured and the outer half escapes (thus, the capture efficiency becomes 0.5). Since a passing object on a hyperbolic orbit has a non-zero value of velocity at infinity, the capture efficiency should be smaller than 0.5 when there is no spin. Therefore, a larger amount of mass escapes from the planet's gravity field than is captured. Neither Dones' formula nor the above formula includes the effect of self-gravity between components of the body. However, the escaping fragments could gravitationally attract other non-captured fragments, and thus, more fragments could escape. The specific gravity of the passing object can be expressed as $E_{\rm grav}=-GM_{\rm body}/r$, where $r$ is the distance from the center of mass of the object. Therefore, the object's gravity becomes stronger as the mass of the object becomes larger, and thus, the effect of self-gravity could play a significant role in the capture efficiency at a larger body.\\

Thus, we model the effect of self-gravity by considering what we call "Hill capture". We calculate the Hill radius of the escaping mass $R_{\rm Hill,esc}=\left( \frac{M_{\rm esc}}{3M_{\rm S}} \right)^{1/3}a_{\rm esc}$, where $a_{\rm esc}$ is the distance between Saturn and the center of mass of the escaping mass. Cells that satisfy $E_{i}<0$ but are within the Hill sphere of escaping fragments are considered as "Hill captured" by the escaping material. Therefore, the mass that is captured by Saturn is the mass that is $E_{i}<0$ and lies outside the Hill sphere of the escaping mass.\\

Figure \ref{effect_of_grav} shows the capture efficiency including Hill capture. Compared to Figure \ref{effect_of_spin}, the self-gravity lowers the capture efficiency, especially at larger pericenter distance. This is because, as the pericenter distance increases, the escaping mass as well as the Hill radius increases, thus Hill capture becomes more efficient. Compared to Dones' formula, in the case of $M_{\rm body}=10^{23}$kg, the captured mass is always lower than Dones' formula. However, when $M_{\rm body}=10^{21}$kg, the capture efficiency could become larger than Dones' formula (see the case of prograde spin in Figure \ref{effect_of_grav} left panel). In this case, it seems that the effect of spin dominates over the effect of self-gravity.\\

The capture efficiency could either be larger or lower than Dones' formula as seen in Figure \ref{effect_of_grav}, depending on spin state and size of the body. Therefore, it is crucial to consider the actual spin state of the body at pericenter. As we show in the next section, the spin at pericenter could be significantly different from the initial spin state at infinity due to the planet tidal torque. In addition, the object is tidally deformed during the close encounter and may be far from spherical at closest approach (see next section). Therefore, even though our above procedures predict better capture efficiency than Dones' formula, direct simulations are necessary in order to obtain a more accurate result. Note that the classical definition of the Roche limit is the critical distance within which a self-gravitational object separates due to tides and this implicitly assumes an infinite duration of the body experiencing the tidal force. However, since the time spent in the Roche limit by a passing object on a hyperbolic orbit is limited, the tidal disruption and fragment capture around a planet only occurs well inside the Roche limit (see also Figures \ref{capture_efficiency_Saturn} and \ref{capture_efficiency_Uranus}).\\

\subsection{Encounter Probability}
In this subsection, we estimate the number of objects that experience close encounters with giant planets within their Roche limit during the LHB. Following the procedure in \cite{Cha09}, the total number of bodies with radius $R_{\rm body}$ that pass between a giant planet's surface and its Roche limit, $N(R_{\rm body})$, can be calculated as
\begin{equation}
   N(R_{\rm body}) = N_{\rm tot}(R_{\rm body}) \times P_{\rm c} (1 + (V_{\rm esc}(a_{\rm R})/V_{\rm inf})^2) (a_{\rm R}^2 - R_{\rm p}^2)
\end{equation}
where, $N_{\rm tot}(R_{\rm body})$ is the total number of bodies with radius $R_{\rm body}$ in the primordial Kuiper belt, $P_{\rm c}$ is the intrinsic impact probability onto giant planets up to the age of the Solar system, $V_{\rm inf}$ is the mean velocity at infinity before the impact, $V_{\rm esc}(a_{\rm R}$) is the escape velocity at the Roche limit for icy material. Using values obtained by re-processing the LHB simulations presented in \cite{Cha09}, we get $P_{\rm c}=4.36 \times 10^{-15}$ km$^{-2}$ and $V_{\rm inf} = 4.22$ km/s for Jupiter, $P_{\rm c}=5.05 \times 10^{-15}$ km$^{-2}$ and $V_{\rm inf}= 3.69$ km/s for Saturn, $P_{\rm c}=1.25 \times 10^{-14}$ km$^{-2}$ and $V_{\rm inf}=2.07$ km/s for Uranus, and $P_{\rm c}=9.66 \times 10^{-15}$ km$^{-2}$ and $V_{\rm inf}=1.70$ km/s for Neptune, respectively\footnote{D. Nesvorn$\acute{\rm y}$ (private communication) also provides similar collision probabilities with slightly ($\sim1.5-2.5$ times) higher values for Jupiter and Saturn.}. Then, using these values, we obtain $N(R_{\rm body})/N_{\rm tot}(R_{\rm body}) \sim$ $7.69 \times 10^{-3}$, $4.10 \times 10^{-3}$, $1.55 \times 10^{-3}$, and $1.74 \times 10^{-3}$ for Jupiter, Saturn, Uranus and Neptune, respectively.

On the other hand, a recent model of primordial Kuiper Belt formation implies the existence of 1000 to 4000 Pluto-sized objects ($M_{\rm body} \sim 10^{22}$kg) and $\sim$ 1000 bodies with a mass twice as massive as Pluto \citep{Nes16}. Thus, it is expected that the giant planets experience at least $\sim$ 8, 4, 2 and 2 close encounters with Pluto-sized bodies within their Roche limit at Jupiter, Saturn, Uranus and Neptune, respectively, and up to a maximum of 32, 16, 8 and 8 encounters. In conclusion, the passing of Pluto-sized objects was not a rare event during the LHB and may have happened several times for each giant planet.

\section{SPH Methods and Models}
Using the smoothed particle hydrodynamics (SPH) method, we perform simulations of large differentiated bodies  passing close to Saturn or Uranus in order to understand the detailed physical process of tidal destruction. The SPH method is a Lagrangian method \citep{Mon92} in which hydrodynamic equations are solved by considering averaged values of particles through kernel-weighted summation. We applied the Tillotson equation of state \citep{Til62} to calculate the pressure from the internal energy and density. For the artificial viscosity, we use a Von Neumann-Richtmyer-type viscosity with the standard parameter sets ($\alpha=1.0$ and $\beta=2.0$). Our numerical code is the same as that used in \cite{Gen15a,Gen15b} and more details are described in their papers.\\

The passing body is assumed to be differentiated with 50w\% silicate core represented by basalt and 50w\% icy mantle represented by water ice. We used the parameter sets of basalt and water ice for Tillotson EOS listed in \cite{Mel89}. The total mass of the body is set to be $M_{\rm body}=10^{21}$ or $10^{23}$kg. We use the total number of $N=100,000$ SPH particles. In our simulations, the gravity of the planet is also taken into account. Saturn and Uranus are represented by a point mass located at the origin of the coordinate system and their masses are $M_{\rm Saturn}=5.69\times10^{26}$kg and $M_{\rm Uranus}=8.68\times10^{25}$kg, respectively. The initial spherical body for our encounter simulations is created by distributing the SPH particles in a 3D lattice (face-centered cubic) and then by performing SPH calculation in isolation until the velocity of particles is much smaller than the escape velocity of the sphere.\\

Initial positions and velocities of all particles follow a hyperbolic orbit around the planet given analytically with initial distance to the planet set to $3.0 \times 10^{5}$ km for Saturn and $1.5 \times 10^{5}$km for Uranus, which are about twice the Roche limit of water ice material. These initial distances are large enough to avoid the tidal effect of the planet (Hyodo \& Ohtsuki 2014). The hyperbolic orbit is entirely determined by the pericenter distance $q$ and the velocity at infinity $v_{\rm inf}$. In our work, we investigate the dependence on pericenter distance but assume a fixed velocity at infinity $v_{\rm inf}=3.0$km/s for Saturn and $v_{\rm inf}=2.0$km/s for Uranus, which are the expected values during the LHB \citep{Cha09}. We note that changing the value of velocity at infinity strongly affects the outcome. Therefore, the outcome is sensitive to the initial states and we will leave this matter for future work. We also investigate the effect of initial spin state of the passing body. The spin-axis of the initial spherical body is assumed to be perpendicular to the orbital plane with either prograde or retrograde spin to the direction of the hyperbolic orbit with the spin period $T_{\rm spin}=8$h.  Such a rotation period is common in the trans-Neptunian belt \citep{Thi14}. We stop our calculation when the mass evolution (e.g. captured mass) in the system achieves a steady-state.

\section{Numerical Results}
\subsection{Case of Homogeneous Body}
First, we consider the simplest case where the passing body is homogeneous and made of  water ice. We perform SPH simulations of close encounters with Saturn with $v_{\rm inf}=3$km/s. Then, we calculate mass that is gravitationally captured around the planet. Figure \ref{homogeneous} shows the captured mass (blue points) against different pericenter distances of the passing object and comparison to Dones' formula (Equation \ref{Dones1991}). Data obtained from SPH simulations are in very good agreement with Dones' formula when the pericenter distance is below $8.3 \times 10^7$m but significant deviations are seen at larger distances. Figures \ref{homo_q7-0} and \ref{homo_q9-1} show snapshots of our simulations for pericenter distances $q=7.0 \times 10^7$m and $q=9.1 \times 10^7$m, respectively. When the pericenter distance is small enough ($q \leq 8.3 \times 10^7$m), the passing body experiences strong tides and is elongated homogeneously into a needle-like structure. Then, due to a small scale gravitational instability, the body splits into several similar-sized clumps (Figure \ref{homo_q7-0}). However, for pericenter distances large enough ($q > 8.3 \times 10^7$m), tides are weaker and the body is no longer stretched, but split into fewer, larger clumps of different sizes  (Figure \ref{homo_q9-1}), which differ from the description in \cite{Don91} that assumes a uniform distribution of the energy in the fragments. Therefore, it seems that Dones' formula is valid when the body experiences an encounter very close to the planet as long as the body is homogeneously elongated. We now return to the case of a differentiated body.\\

\subsection{Captured Mass around Giant Planets}
Figure \ref{capture_efficiency_Saturn} shows the capture efficiency around Saturn obtained from our SPH simulations (dots) for incoming differentiated bodies with masses $M_{\rm body}= 10^{21}$kg (left panels) and  $10^{23}$kg  (right panels) as well as  Dones' formula (dashed line). The different coloured dots represent the silicate mass fraction of the total captured mass ($M_{\rm sil,cap}/M_{\rm cap,tot}$). The light blue regions in the panels represent areas within the radius of the planet, and thus such encounters never actually occur (here, in order to understand detailed physics of tidal disruption, we consider these cases too).\\

Like in the case of a homogeneous body, numerical results deviate from Dones' formula. The slight jump observed in the capture efficiency in the case of prograde and no-spin cases (e.g. between $5.6 \times 10^7$m and $6.3 \times 10^7$m in the case of $M_{\rm body}=10^{23}$kg with no initial spin) corresponds to the point where the core also starts to be tidally destroyed (see also Figures \ref{differ_T-8h_q5-6} and \ref{differ_Tinf_q7-0}). As discussed in the previous section, in the case of a light object (case of $M_{\rm body}=10^{21}$kg, left panels), the prograde spin effects dominate over the self-gravity effect (Hill capture) and thus more mass than that predicted by Dones' formula is captured. Conversely, in the case of a heavier object (case of $M_{\rm body}=10^{23}$kg, right panels), the effect of self-gravity (Hill capture) dominates and more mass escapes (thus, the capture efficiency decreases).\\

On the other hand, in the case of retrograde spins (both left and right bottom panels in Figure \ref{capture_efficiency_Saturn}), neither Dones' formula nor our semi-analytical model is applicable. In addition, at small pericenter distances (below $5 \times 10^7$m), capture efficiency becomes constant at around 0.08. During such close encounters, an object with an initial  retrograde is spun up in the prograde direction by tidal forces and experiences a major and continuous complex change of its shape. We also confirm an increase in the internal energy of the constituent SPH particles during the encounter. Thereby the body heats up and mostly ends up splitting into two large objects (Figure \ref{differ_T8h_q4-2}), but the increase of internal energy is not enough to melt the icy material. These effects are not included either in Dones' formula or our semi-analytical model. However, in the prograde and no-spin cases, even though our semi-analytical model cannot reproduce numerical results, it is still better than Dones' formula as it includes spin effects and self-gravity (Hill capture).\\

Figure \ref{capture_efficiency_Uranus} is the same as Figure \ref{capture_efficiency_Saturn} but for the case of Uranus. General trends of the capture efficiency are similar to the case of Saturn. However, since the density of Uranus is larger than that of Saturn, the passing object can physically pass deeper inside the potential field of the planet. Thus, the objects can be tidally destroyed more significantly than in the case of Saturn without directly colliding with the planet (outside the light blue region in the figure), resulting in a higher silicate fraction of the captured mass.

\subsection{Orbits of Captured Fragments and Resulting Ring Mass}
In this section, we discuss the orbits of captured fragments. During the encounter, the orbital kinematic energy is redistributed among the fragments and thus it changes their semi-major axes. Then, due to the conservation of the orbital angular momentum, captured fragments have smaller eccentricities and escaping fragments have larger eccentricities than the initial body. The eccentricity distribution depends on the initial size of the body and so the distribution range is wider for initially larger bodies as they can be more elongated by tides. Considering the conservation of the specific orbital angular momentum of a passing object $J_{\rm 0}=\sqrt{M_{\rm 0}Ga(1-e^2)}$, we can derive the relationship between semi-major axis $a$ and eccentricity $e$ as 
\begin{equation}
	e=\sqrt{1-\frac{J_0^2}{M_0Ga}}.
\label{ea_capture_analytic}
\end{equation}
Figure \ref{ea_fragments} shows eccentricities of the captured fragments (black dots) as well as the relation derived from Equation (\ref{ea_capture_analytic}) (case of $M_{\rm body}=1 \times 10^{21}$kg (left panel) and $10^{23}$kg (right panel), prograde spin with $T_{\rm spin}=8$h, $q=7.0 \times 10^{7}$m and $v_{\rm inf}=3.0$km/s around Saturn). The slight increase of the eccentricities of SPH data from the analytical estimation is due to the transfer of the orbital angular momentum to the spin angular momentum of the body during the encounter.\\

Thus, captured fragments have large eccentricities around $0.9$ (case of $10^{23}$kg body) and $0.98$ (case of $10^{21}$kg body).  Our SPH simulations are stopped before the captured fragments reach their apocenter. However, if their apocenter distance is larger than the planet's Hill radius, or their pericenter distance is smaller than the planet's radius, then they will either escape from or collide with the planet, so that they cannot be incorporated into the rings. Figures \ref{pericenter_apocenter_Saturn} and \ref{pericenter_apocenter_Uranus} show pericenter distances and apocenter distances of captured fragments in our SPH simulations (note that, in the figures, different coloured dots represent initial different pericenter distances of the passing objects). In the panels, the light red area corresponds to the region where the fragment's apocenter is larger than the Hill radius of the planet and the blue area is where their pericenter is smaller than the planet's radius. In Figures \ref{capture_efficiency_Saturn} and \ref{capture_efficiency_Uranus}, the mass of the captured fragments outside the red region is summed up and shown.\\

Now, we discuss how much mass can be incorporated into the rings. First, we calculate equivalent circular orbital radius of the captured fragments using their center of mass position and velocities as 
\begin{equation}
	a_{\rm eq}	=a_{\rm sim}\left(1-e_{\rm sim}^2 \right)
\label{aeq}
 \end{equation}
where $a_{\rm sim}$ and $e_{\rm sim}$ are semi-major axis and eccentricity obtained from SPH simulations. Then, any fragments that satisfy 
 \begin{equation}
 	a_{\rm eq}>R_{\rm pla}
\end{equation}
as well as  
\begin{equation}
	r_{\rm peri} > R_{\rm pla} \hspace{1.5em} {\rm and } \hspace{1.5em}  r_{\rm apo} < R_{\rm Hill}
\end{equation}
where $R_{\rm pla}$ is the radius of the planet, $r_{\rm peri}$ is the pericenter distance, $r_{\rm apo}$ is the apocenter distance and $R_{\rm Hill}$ is the Hill radius of the planet, are counted as ring mass; this includes all those in the white region in Figures \ref{pericenter_apocenter_Saturn} and \ref{pericenter_apocenter_Uranus}. \\
 
Figures \ref{Ring_mass_Saturn} and \ref{Ring_mass_Uranus} show the mass that can be incorporated into the rings. As we described in Section 4.2, more mass can be captured by the planet with decreasing pericenter distance of the passing body. On the other hand, when the fragment's pericenter distance decreases, the fraction of the fragments that collide with the planet increases (see Figures \ref{pericenter_apocenter_Saturn} and \ref{pericenter_apocenter_Uranus}). Therefore a smaller ring mass is obtained in some cases as the pericenter distance of the passing object becomes smaller (Figures \ref{Ring_mass_Saturn} and \ref{Ring_mass_Uranus}). In both Saturn and Uranus cases, the ring mass is about 0.1-10\% of the initial mass of the passing object. However, since the density of Uranus is higher than that of Saturn, an object can penetrate deeper inside Uranus' Roche limit than Saturn's and experience more significant tidal destruction. Therefore, in the case of Uranus, the resulting silicate fraction in the ring mass can be much higher than in the case of Saturn. This argument holds for Neptune as well, as it is also denser than Saturn. This could explain the observation that Uranus and Neptune rings are darker than Saturn's rings as they are likely to have a higher rock fraction \citep{Tis13}. The ratio of $a_{\rm eq}$ to the initial pericenter distance of captured fragments $q$ is $a_{\rm eq}/q=(1+e)$. Since fragments' eccentricities are mostly larger than $0.9$ after the tidal destruction, the ratio is about $2$. Therefore, from the results shown in Figures \ref{pericenter_apocenter_Saturn} and \ref{pericenter_apocenter_Uranus}, the radial locations of circular rings are expected to be between $120-180 \times 10^{6}$m ($0.86-1.29R_{\rm Roche}$) in the case of Saturn and $50-90 \times 10^{6}$m ($0.71-1.29R_{\rm Roche}$) in the case of Uranus, where $R_{\rm Roche}$ is the planet's Roche limit.

\section{Long-term Evolution of the Captured Fragments}
In the previous sections, by using SPH simulations, we have shown that during a close encounter of a passing object with Saturn or Uranus, some fragments are captured. However, they are still in the form of large clumps ($m \sim 10^{18-19}$kg in the case of $M_{\rm body}=10^{21}$kg and $m \sim 10^{20-21}$kg in the case of $M_{\rm body}=10^{23}$kg) and are on highly eccentric orbits ($e \sim 0.9-0.98$). Therefore, it is necessary to investigate their longer-term evolution and see if they can settle into a collection of small particles on nearly circular orbits as seen in current ring systems. In this section, we examine the long-term evolutions of captured fragments by using N-body simulations as well as analytical estimates. As an example, we use the case of SPH data obtained using a passing body ($M_{\rm body}=10^{21}$kg) on a prograde spin ($T_{\rm spin}=8$h), with $q=7.0\times10^{7}$m and $v_{\rm inf}=3$km/s around Saturn.\\

\subsection{N-body Methods and Initial Conditions}
Using the data obtained from our SPH simulations, we perform N-body simulations on the longer-term evolution of their successive orbits around the giant planets. Our N-body code is essentially the same as that used in \cite{Hyo15a}, but we now include the effect of a planet's oblate potential up to the factor $J_4$ for Saturn as given by \citep[e.g.][]{Had99}
\begin{equation}
	J_2 = 0.016298, \hspace{2em} J_4 = -0.000915.
\end{equation} 
The equation of motions are 
\begin{eqnarray}
	\ddot{x}_i = & -GM_0 \left\{  \displaystyle \frac{x_i}{r_i^3} \left( 1 - J_2\Psi_{i2} -J_4 \Psi_{i4} \right) - \sum_{j \neq i}  m_j  \displaystyle \frac{x_j - x_i}{r_{ij}^3} \right\} \\	
	\ddot{y}_i = & -GM_0 \left\{  \displaystyle \frac{y_i}{r_i^3} \left( 1 - J_2\Psi_{i2} -J_4 \Psi_{i4} \right) - \sum_{j \neq i}  m_j  \displaystyle \frac{y_j - y_i}{r_{ij}^3} \right\} \\
	\ddot{z}_i = & -GM_0 \left\{  \displaystyle \frac{z_i}{r_i^3} \left( 1 - J_2\Psi_{i2} -J_4 \Psi_{i4} + J_2\Phi_{i2} + J_4\Phi_{i4} \right) - \sum_{j \neq i} m_j  \displaystyle \frac{z_j - z_i}{r_{ij}^3} \right\} 
 \end{eqnarray}
where $\bm{r}_{i}=\left( x_i,y_i,z_i \right)$ is the position vector of the $i$-th particle, and 
\begin{eqnarray}
	\Psi_{i2} = \displaystyle \frac{R_{\rm pla}^2}{r_i^2} P'_3 \left( \displaystyle \frac{z_i}{r_i} \right), \hspace{2em} \Psi_{i4} &= \displaystyle \frac{R_{\rm pla}^4}{r_i^4} P'_5 \left( \displaystyle \frac{z_i}{r_i} \right)\\
	\Phi_{i2} = 3\displaystyle \frac{R_{\rm pla}^2}{r_i^2}, \hspace{2em} \Phi_{i4} &= \displaystyle \frac{R_{\rm pla}^4}{r_i^4} Q_4 \left(\displaystyle \frac{z_i}{r_i} \right)
\end{eqnarray} 
with
\begin{eqnarray}
	P'_3(x) = \displaystyle \frac{15}{2}x^2 - \displaystyle \frac{3}{2}, \hspace{2em} P'_5(x) = \displaystyle \frac{315}{8}x^4  - \displaystyle \frac{105}{4}x^2 + \displaystyle \frac{15}{8}\\
	Q_4(x) = \displaystyle \frac{35}{2}x^2 - \displaystyle \frac{15}{2}.
\end{eqnarray} 
\\

In our N-body model, particles are considered to be smooth spheres and collisions between them are solved by using a hard-sphere model. When two particles collide, we compute the velocity change using a normal coefficient of restitution $\epsilon_{n}=0.1$. Any particles that hit the planet are removed. Initial fragments are modelled as hexagonally close-packed spherical aggregates consisting of same-sized particles (radius $\sim 26$km) with density $\rho=1300$kg/m$^3$ (Figure \ref{Nbody_initial}). Here we focus on the initial dynamical evolutions of particles but do not investigate the long-term flattening because of computer limitations. The total number of particles is $N\sim1000$. Initial positions and velocities of the center of mass of the fragments are the same as those obtained from SPH simulations with no initial spin (Figure \ref{Nbody_initial}). 

\subsection{Tidal Destruction of Captured Fragments}
After running the N-body simulation we find that aggregates are progressively tidally destroyed after a few orbits ($\sim$ few years) because their pericenters are well inside the planet's Roche limit (see Figures \ref{pericenter_apocenter_Saturn} and \ref{pericenter_apocenter_Uranus}), and thus form an eccentric ringlet-like structure with a large eccentricity ($e\sim 0.98$). As particles have slightly different semi-major axes (see purple dots in Figure \ref{ea_fragments}), when their separation is $\sim f r_{\rm p}$ ($f$ is a factor of the order of unity and $r_{\rm p}$ is the particle radius) from the closest particles, neglecting the effect of oblateness potential of the planet, their collision timescale can be estimated using their synodic period as 
\begin{equation}
	\tau_{\rm col,syn} \sim T_{\rm syn} = 2\pi a /\left(  \frac{3}{2} fr_{p} \Omega \right)
\end{equation}
where $a$ is the semi-major axis and $\Omega$ is the orbital frequency.  Using $a=5 \times 10^9$ m and $r_{\rm p}=26$ km which are similar to the values in the N-body simulation, we get $\tau_{\rm coll, syn} \sim 46,000f^{-1}$ years. In the next subsection, we will compare this collision timescale with the precession timescale.\\

\subsection{Precession of Captured Particles}
The precession rate of the argument of pericenter $\omega$ and the longitude of ascending node $\Omega$ due to the $J_2$ term can be described as \citep{Kau66}
\begin{equation}
	\dot{\omega} = \frac{3n}{\left( 1-e^2 \right)^2} \left( \frac{R_{\rm pla}}{a} \right)^2 \left( 1-\frac{5}{4}\sin^2(i) \right)J_2
\end{equation}
\begin{equation}
	\dot{\Omega} =  -\frac{3n\cos(i)}{2\left( 1-e^2 \right)^2}J_2
\end{equation}
where $n$ is orbital mean motion and $i$ is the inclination from the planet's equatorial plane. The precession timescale of the argument of pericenter and of the longitude of ascending node are  
\begin{equation}
 	\tau_{\rm \omega,pre}=2\pi/\dot{\omega}
\end{equation}
\begin{equation}
 	\tau_{\rm \Omega,pre}=2\pi/\dot{\Omega}.
\end{equation}

For Saturn ($R_{\rm Saturn}=6.0 \times 10^7$m) with $a=5\times10^9$m, $e=0.98$ and $i=10, 45$ and $80$ degrees, for example, we have $\tau_{\rm \omega,pre} \sim 80, 210$ and $380$ years, and $ \tau_{\rm \Omega,pre} \sim 160, 230$ and $930$ years, respectively. Therefore, 
\begin{equation}
	\tau_{\rm \omega,pre}, \tau_{\rm \Omega,pre} \ll \tau_{\rm col,syn}, 
\end{equation}
and thus precession dominates over the collisional evolution after initial aggregates are tidally destroyed.\\

Thus, neglecting collision, we perform an N-body simulation under the effect of the $J_2$ and $J_4$ terms assuming the radius of the constituent particles of aggregates $r_{\rm p}=260$km. The initial orbital plane is assumed to be inclined by $45$ degrees with respect to the planet's equatorial plane. Figure \ref{Nbody_no_collision} shows the longitudes of ascending node (left panel) and the arguments of pericenter (right panel) at different times. At smaller semi-major axes, they are rapidly  randomised. The timescale of the precession is around hundreds years consistent with the above discussion, and eventually the system will form a torus-like structure.\\

\subsection{Long-term Collisional Evolution}
In this section, we discuss what kind of collision could occur as well as its timescale after particles form a torus-like structure. In the particle-in-a-box approximation, the collision timescale can be written as 
\begin{equation}
	\tau_{\rm col, ran} \sim \frac{1}{n \sigma_{\rm col} v_{\rm rel}}
	\label{timescale}
\end{equation}
where $n$ is the number density of particles, $\sigma_{\rm col}$ is the collision cross section and $v_{\rm rel}$ is the relative velocity. Here, particles still have large eccentricities ($e \sim 0.9-0.99$) with randomised arguments of pericenter and longitudes of ascending node. Therefore, orbital crossing occurs and  $v_{\rm rel}$ can be given by
\begin{equation}
	v_{\rm rel} \sim  v_{\rm Kep}\left(  e^2 + \sin(i)^2 \right)^{1/2}
\end{equation}
where $v_{\rm Kep}$ is the Keplerian velocity. Since eccentricity $e \sim 1$ and inclination, for example, $i = 45$ degrees, the relative velocity becomes about Keplerian velocity.  We have confirmed such collisions occur by using N-body simulations. The collision cross section is written as
\begin{equation}
	\sigma_{\rm col} = 4 \pi r_{\rm p}^2. 
\label{sigma}
\end{equation}
Here we neglect the gravitational focusing term because, when particles are closer or within the Roche limit, gravitational attraction between particles becomes negligible \citep{Oht93, Hyo14}. The volume number density of particles is (Appendix) :
\begin{equation}
	n= \frac{ N_{\rm tot} P(r) P(\psi)}{2 \pi r^2} 
\label{number_density}
\end{equation}
where $P(r)$ and $P(\psi)$ are the probability of finding a particle at radial distance $r$ and an angle $\psi$ from the equatorial plane of the planet and are written as  
\begin{equation}
	P(r)=\frac{r}{\pi a \sqrt{a^2e^2-(a-r)^2}}
\end{equation}
and 
\begin{equation}
	P(\psi)=\frac{|\cos(\psi)|}{\pi \sqrt{ \sin(i)^2 - \sin(\psi)^2}}, 
\end{equation}
respectively.

Using $v_{\rm rel} \sim v_{\rm Kep}$ as well as Equations (\ref{sigma}) and (\ref{number_density}), we can get the collision timescale (Figure \ref{tau_coll_Seb}).  Particles are more prone to collide when closer to the planet and collision velocity could be larger than a few km/s. Such a high collision velocity would be destructive enough and could generate smaller fragments as seen in current ring systems \citep{Ara99,Ste12} or might vaporise them \citep{Kra11}. However, including such physical effects is beyond the scope of our present work and we will leave them to future works. Collisional destruction as well as inelastic bouncing between particles is expected to damp their eccentricities and inclinations, and thus they would eventually form thin equatorial circular rings \citep[see also][]{Mor12} made of small particles.\\

\subsection{Effect of a Primordial Satellite}
In this subsection, we discuss the potential effect of a primordial satellite on the dynamical evolution of the captured mass. Saturn's largest satellite Titan is thought to have formed by accretion in the protoplanetary gas/dust disk  \citep{Can09,Est09}. Thus, at the time of LHB, Titan would have already been in orbit around Saturn. Since the orbits of captured mass are highly eccentric (Section 5.2) - apocenter (or pericenter) distance is larger (or smaller) than that of Titan - the captured mass could be removed either by gravitational scattering or accretion by Titan. Here, we consider only Titan as a primordial satellite. If Dione or Rhea was also already present, the removal could be enhanced.\\

Here we perform additional simulations including Titan, assuming its equatorial orbit has the current eccentricity of $0.0288$. The captured fragments are represented by test particles using the SPH data obtained for $M_{\rm body} = 10^{21}$kg with a prograde spin ($T_{\rm spin} = 8$h), $q = 7.0 \times 10^{7}$m and $v_{\rm inf} = 3$km/s. Initial orbital plane of the captured fragments is set to be 0, 25 or 45 degrees with respect to the equatorial plane. Any particles that collide with either Titan or Saturn, or that go beyond 10 times of Saturn's Hill radius are removed.\\

Figure \ref{effect_of_titan} shows the results of N-body simulations. Only when the initial orbital plane of captured fragments is the same as that of Titan ($0$ degrees), their orbits cross that of Titan and the effect of accretion by Titan is significant. However, when orbits of captured fragments are inclined ($25$ or $45$ degrees) with respect to the equatorial plane, direct collision with Titan rarely happens. In contrast, collision with Saturn as a result of gravitational scattering by Titan is common in all the cases (without Titan, we confirm that accretion by Saturn does not occur). The mass that escapes from Saturn's Hill sphere is negligible (few percent or less). The timescale and efficiency of removal of captured mass due to the effect of Titan depends on the initial inclination of captured mass such that those fragments on the more inclined orbits are harder to remove. Note that, in Figure \ref{effect_of_titan}, the total removed mass keeps increasing over time. However, in the real system (including collisional dumping), the collision would damp the system and further removal is expected to cease at some point.\\

In the case of 0 degrees, about 50\% of the captured mass is removed after 100 years (Figure \ref{effect_of_titan}). The precession timescale and successive collision timescale are comparable to or larger than 100 years (see Figure \ref{tau_coll_Seb}). Thus, a large portion of captured fragments could be removed before they settle in thin rings in this case. However, when the initial orbital plane is inclined by 25 or 45 degrees, only about 40 or 20\% is removed even after 1000 years. The timescale of precession and successive collisional damping is shorter than or comparable to the above timescale for the mass removal (see Figure  \ref{tau_coll_Seb}). Thus, a large portion of captured fragments is expected to survive even under the existence of Titan, unless the fragments happened to be distributed near Titan's orbit with very low inclinations.\\

Now, one may wonder whether a single collision is enough to damp the eccentric orbit inside that of Titan to avoid successive close encounters. Here we simply estimate the change of orbital elements after one collision around Saturn. Assuming two particles with mass $m_{i=1,2}$ collide with each other, the orbit of colliding particles is changed while conserving energy and angular momentum as 
\begin{eqnarray}
	 -\frac{GM_{\rm Sat}m_{1}}{2a_{\rm 1,new}} - \frac{GM_{\rm Sat}m_{2}}{2a_{\rm 2,new}}  =  -\frac{GM_{\rm Sat}m_{1}}{2a_{1}} - \frac{GM_{\rm Sat}m_{2}}{2a_{2}} - \Delta E_{\rm col}\\
	 m_{1} \sqrt{GM_{\rm Sat}a_{\rm 1,new}(1-e^{2}_{\rm 1,new})} + m_{2} \sqrt{GM_{\rm Sat}a_{\rm 2,new}(1-e^{2}_{\rm 2,new})} =\\
	 m_{1} \sqrt{GM_{\rm Sat}a_{\rm 1}(1-e^{2}_{\rm 1})} + m_{2} \sqrt{GM_{\rm Sat}a_{\rm 2}(1-e^{2}_{\rm 2})}
\end{eqnarray}
where $a_{i=1,2,\rm new}$, $a_{i=1,2}$ and $e_{i=1.2,\rm new}$, $e_{i=1,2}$ are semi-major axes and eccentricities of the colliding particles after and before collision. $ \Delta E_{\rm col} = \frac{1}{2} \left( 1 - \epsilon_{\rm n}^2 \right) \mu V^{2}_{\rm imp}$ is the dissipated energy during the collision with the normal coefficient of restitution $\epsilon_{\rm n}=0.1$ and the reduced mass $\mu=m_{1}m_{2}/\left(m_{1} + m_{2} \right)$. The impact velocity is about the Keplerian velocity at the radial distance $r$ from Saturn as $V_{\rm imp} \sim V_{\rm K} = \sqrt{GM_{\rm Sat} \left( 2/r - 1/a \right)}$ because the orbit of particles crosses as a result of precession. Then, assuming the simplest case with $m \sim m_{1} \sim m_{2}$, $a \sim a_{1} \sim a_{2}$, $e \sim e_{1} \sim e_{2}$ and $a_{\rm new} \sim a_{\rm new,1} \sim a_{\rm new,2}$, $e_{\rm new} \sim e_{\rm new,1} \sim e_{\rm new,2}$, new orbital elements after collision are derived as a function of those of initial values as 
\begin{eqnarray}
	&a_{\rm new} = \left( 1/a + \Delta E_{\rm col}/(GM_{\rm Sat}m) \right)^{-1}\\
	&e_{\rm new} = \sqrt{1 - a/a_{\rm new}(1-e^{2}}).
\end{eqnarray}
With the initial eccentricity and semi-major axis of two particles, for example, $e=0.98$, and $a=5 \times 10^{9}$m, and assuming that the collision takes place at the Roche limit ($\sim135,000$km) because the collision probability is higher closer to the planet (Figure \ref{tau_coll_Seb}), we get $a_{\rm new} \sim 2.6 \times 10^{8}$m and $e_{\rm new} \sim 0.49$. This means that the apocenter of the new orbit after the collision is well inside the orbit of Titan ($ \sim1.2 \times 10^{9}$m). Thus, damping due to a single collision can sufficiently shrink the initial orbit of captured fragments so that they settle inside the orbit of Titan, avoiding the second encounter with Titan after the collision. Of course, this is a rather simple estimate and neglects detailed collisional processes such as destruction, compaction or vaporisation. Thus, direct numerical simulations are clearly needed in future work.

\section{Discussion \& Conclusions}
The origin of rings around planets is still debated. In this work, we investigate the possibility that a single close encounter of a large differentiated body may form rings around giant planets (Figure \ref{summary}). We perform two different direct simulations (SPH simulations and N-body simulations) in order to understand the physics of the tidal disruption of a passing large differentiated primordial KBO and long-term evolution of the resultant captured fragments. 

At the time of LHB, we assume that giant planets could suffer a significant number of encounters with primordial KBOs \citep[see also][]{Cha09}. In the previous works, such a process was investigated using a simplified analytical formula for the capture efficiency of mass around a planet \citep{Don91, Cha09}. However, the physical model in these works only considered ballistic orbits of the constituent particles and neglected the self-gravity and spin state of the passing body. 

Recently, \cite{Nes16} have argued that the primordial Kuiper belt should have had $1000$ to $4000$ Pluto-sized bodies ($M_{\rm body} \sim 10^{22}$kg) in order to explain the observed abundance of objects on non-resonant orbits in the present Kuiper belt. Computation of encounter probabilities (see Section 2.3) suggests that all giant planets may have experienced a few to several tens of encounters inside their Roche limit with Pluto-sized objects during the LHB. Thus, this big reservoir of impactors might have delivered significant mass to the giant planet system through tidal disruption.

Using semi-analytical arguments and SPH simulations, we first investigate the detailed physical process of tidal disruption of a passing large differentiated object. Our semi-analytical model shows significant effects of the self-gravity on the captured mass and we find that Dones' (1991) formula for the capture efficiency is not always consistent with our simulation results. When the object is large enough ($M_{\rm body}=10^{23}$kg), the capture efficiency tends to be smaller than Dones' formula regardless of the spin state of the object. In contrast, when the object is smaller ($M_{\rm body}=10^{21}$kg), the capture efficiency can be larger and smaller than Dones' formula depending on the direction of the spin of the object. Such deviation from Dones' formula becomes larger as the pericenter distance becomes larger. We also investigated the effect of an initial spin of the passing body  and found that destruction becomes more significant and capture efficiency becomes larger when the body has an initial prograde spin with respect to the direction of the encounter, while the capture efficiency becomes smaller when the body has an initial retrograde spin. SPH simulations show that, during a close encounter, tidal forces spin up the body in the prograde direction. Therefore, a body with an initial prograde spin is more easily and efficiently deformed (smoothly elongated) and destroyed than a body with an initial retrograde spin. If the body's pericenter is well inside the planet's Roche limit and above the planet's radius, then the body is disrupted and about 0.1-10\% of its mass is captured and expected to end up in a ring-structure. Assuming an average capture probability of about 1\% and encounter rates reported in Section 2.3, the total mass delivered to giant planets through the tidal break-up of objects with the mass of $M_{\rm body}=10^{22}$kg may range from 8, 4, 2 and $2 \times 10^{20}$kg (for Jupiter, Saturn, Uranus and Neptune, respectively) up to 32, 16, 8 and 8 $\times 10^{20}$kg.

Close encounters of small bodies with planets is a natural consequence of the LHB. As was also discussed in \cite{Cha09}, tidal disruption of passing objects could form massive rings around all giant planets. In addition, recent works have shown that the inner regular satellites of Saturn, Neptune and Uranus could form by spreading of ancient massive rings \citep{Cha10,Cri12,Hyo15a}, and narrow rings such as Saturn's F ring and Uranian $\epsilon$ ring with their shepherding satellites could be the natural consequence at the last stage of such ring spreading across the Roche Limit \citep{Hyo15b}. During a close encounter of a $M_{\rm body} = 10^{21}$kg passing object with Saturn, our numerical simulations show that only enough mass to explain Saturn's current ring ($ \sim 10^{19}$kg) can be embedded. On the other hand, when the object has the mass $M_{\rm body} = 10^{23}$kg, enough mass not only for the current rings but also for the total mass of its regular satellites (up to and including Rhea) can be embedded.  In addition, we find a small fraction of silicate material in the captured mass because the object's mantle is preferentially disrupted. In the case of Uranus, the captured mass could explain only the mass of the current rings (mass of $10^{15}-10^{16}$ kg \citep{Fre91}), assuming an encounter with a body with mass $M_{\rm body} = 10^{21}$kg. In contrast, ring mass as well as the mass of all satellites up to Oberon ($\sim 10^{22}$kg) can be explained by $M_{\rm body} = 10^{23}$kg body. Furthermore, in the case of Uranus, due to its higher density, the width between the planet's surface and its Roche limit is larger than in the case of Saturn. Thus, a body can pass deeper potential field of the planet. As a result, the tidal destruction could be significant enough to disrupt not only the body's icy mantle but also its silicate core, and thus silicate components can be more efficiently captured than in the case of Saturn. This would also be applicable to Neptune since it is also denser than Saturn. Therefore, this could explain the fact that the rings of Uranus and that of Neptune are darker than that of Saturn \citep{Tis13} \\

Soon after the capture, the fragments are in the form of big chunks ($m \sim 10^{18-19}$kg in the case of $M_{\rm body} = 10^{21}$kg and $m \sim 10^{20-21}$kg in the case of $M_{\rm body} = 10^{23}$kg) and they have very large eccentricities ($e \sim 0.9-0.98$). Therefore, their long-term evolution needs to be investigated to see if the system becomes circular equatorial rings as currently seen in the ring systems around planets. Thus, using the data obtained from SPH simulations, we perform N-body simulations of the captured fragments including the oblate potential of the planets. Since their pericenter distances are deep inside the planet's Roche limit, after several Keplerian orbits ($\sim$ year), the fragments are further tidally destroyed and form a ringlet-like structure which still has a large eccentricity on the same plane. However, they would still be large particles ($\sim$km to 100km) assuming their physical tensile strength for pure ice at low temperatures \citep{Har78}.\\

Due to orbital precession, these individual particles form a torus-like structure ($\sim$ hundreds-to-thousand years). In this configuration, orbital crossing between particles is enhanced and particles could experience high velocity collisions ($v_{\rm coll} \sim$ few km/s). Tidal disruption itself can only form km sized fragments. However, such highly energetic collisions would be catastrophic enough to grind them down to centimetre-to-meter sized particles that are currently seen in Saturn's main rings or some vaporisation of particles might occur. Eventually, destruction or inelastic collisions leads to energy dissipation of random velocities and the cloud of debris would form a thin equatorial ring inside the planet's Roche limit. Note that direct simulations are still necessary, including such destruction and vaporisation, in order to understand more details about the final state of the ring systems. However, such complicated physics is beyond the scope of this paper and we will leave it to future works.\\

The above scenario seems to be very consistent with the satellite formation picture depicted in \cite{Cha11}, where an initial massive ring system spreads and gives birth to satellites at the Roche Limit. The silicate content of Saturn's satellites may come from big chunks of silicate initially implanted at random in the primordial ring system. The tidal disruption scenario may naturally form such big chunks of silicate. These chunks may also constitute the "silicate shards" of material thought to be embedded in the core of propeller structures observed in rings by Cassini \citep{Tis06, Por07}.\\

In conclusion, the tidal disruption of a passing large differentiated KBO could form massive rings around giant planets and could explain the compositional differences between the rings of Saturn and those of Uranus (and possibly those of Neptune too). However, the major caveat of this model is the question of angular momentum \citep{Esp91}. Rings and satellite systems of the four giant planets rotate in the prograde direction. However, encounters of KBOs with giant planets happened from random directions, so that the net total angular momentum should be close to zero in the case of a very large number of encounters (relevant for comet-sized objects). However, the number of encounters with Pluto-sized objects is quite small. So small number statistics may apply so that, by chance, only a couple of encounters may dominate the total angular momentum budget for a given planet. Therefore, having a system rotating in the prograde direction may mostly be a matter of chance. Whereas this explanation may hold for one planet, it is hard to believe that the same encounter's history happened for all giant planets. We think this problem might be solved by considering the effect of tides on retrograde objects. All fragments on retrograde orbits will experience a stong negative tidal torque from the giant planet leading to their orbital decay and crash into the planet. The elimination timescale depends strongly on the size of the fragments, but such information is not accurately provided by the current SPH simulations. However, if the tidal decay time is smaller than the collisional flattening time, this would imply that all retrograde material could be removed before the disk is formed in the equatorial plane. This process is far beyond the scope of our paper and thus we leave this matter to future work.

One of the most striking features of giant planet ring systems is their diversity. Whereas Saturn's rings are massive, those of Jupiter, Uranus and Neptune are much less massive and show different structures. Our model suggests that all giant planets could have initially had massive rings as in \cite{Cri12}. However, they could have different subsequent evolution: as noted in \cite{Cha09b}, Saturn's ring is the only system with a synchronous orbit inside the Roche limit, whereas the other planets have opposite relative locations. This means that any satellite forming at the edge of  the Roche limit of Jupiter, Uranus or Neptune should tidally decay inside the ring system and may push the rings inwards due to resonant interactions, leading to a faster removal of the ring system. This might explain the diversity of today's ring system even though they have all started from massive rings.  However, the satellite might be tidally destroyed once it enters deep inside the Roche limit and additional ring material could be supplied. This process is also far beyond the scope of our paper and thus we leave this matter to future work. 

\acknowledgments
We thank D. Nesvorn$\acute{\rm y}$ for useful discussions. R.H. is grateful to Shigeru Ida and Masahiko Arakawa for discussion. This work was supported by JSPS Grants-in-Aid for JSPS Fellows (15J02110) and by Scientific Research B (22340125 and 15H03716). Part of the numerical simulations were performed using the GRAPE system at the Center for Computational Astrophysics of the National Astronomical Observatory of Japan. We acknowledge the financial support of the UnivEarthS Labex programme at Sorbonne Paris Cit{\'e} (ANR-10-LABX-0023 and ANR-11-IDEX-0005-02). This work was also supported by Universit{\'e} Paris Diderot and by a Campus Spatial grant. S{\'e}bastien Charnoz thanks the IUF (Institut Universitaire de France) for financial support.


\appendix
\section{Number density of particles}
Here we estimate the probability of particle positions on their orbits around a planet. The classical Kepler's equation is 
\begin{equation}
	M=E-\sin E
\end{equation}
where $M$ is the mean anomaly and $E$ is the eccentric anomaly. The relationship between the radial distance $r$ from the planet and $E$ can be written as

\begin{equation}
	r=a\left( 1-e\cos E \right)
\end{equation}
where $a$ and $e$ are semi-major axis and eccentricity of a particle, respectively. Then, the probability density of a particle to be at the distance $r$ is written as 

\begin{equation}
	P(r)=2P(E) \frac{dE}{dr} =2 P(M)  \frac{dM}{dE}  \frac{dE}{dr} 
\end{equation}
where $P(E)$ and $P(M)$ are the probabilities of finding a particle at $E$ and $M$, respectively. Note that the factor $2$ comes from the symmetricity of the orbit on $E$ or $M$ to $r$. Since $P(M)=1/2\pi$, we get 

\begin{equation}
	P(r)=\frac{r}{\pi a \sqrt{a^2e^2-(a-r)^2}}
\end{equation}

Next, we try to estimate the probability of the vertical direction ($z$ direction which is perpendicular to the equatorial plane of the planet). We define an angle of $\psi$ from the equatorial plane as 

\begin{equation}
	\psi = \pm \sin^{-1}(z/r)
\end{equation}
The $z$ position is related to particle orbital elements as 

\begin{equation}
	z=r \sin (\omega + \Omega ) \sin i
\end{equation}
where $\omega$ is the argument of pericenter and $\Omega$ is the longitude of ascending node. Using the longitude of pericenter  $\varpi=\omega + \Omega$, we can write the probability of particle at the angle $\psi$ as 

\begin{equation}
	P({\psi}) = 2P(z)\frac{dz}{d\psi} = 2P(\varpi) \frac{d\varpi}{dz} \frac{dz}{d\psi}
\end{equation}
where $P(z)$ and $P(\varpi)$ are the probabilities of finding a particle at $z$ and $\varpi$, respectively, and the factor 2 comes from the symmetricity of the orbit on $\varpi$ or $\psi$ on $z$. Since $P(\varpi)=1/2\pi$, we get 

\begin{equation}
	P(\psi)=\frac{|\cos(\psi)|}{\pi \sqrt{ \sin(i)^2 - \sin(\psi)^2}}.
\end{equation}

\clearpage
 \makeatletter
    \renewcommand{\thefigure}{%
    \arabic{figure}}
 \makeatother
  
\begin{figure}
  \plotone{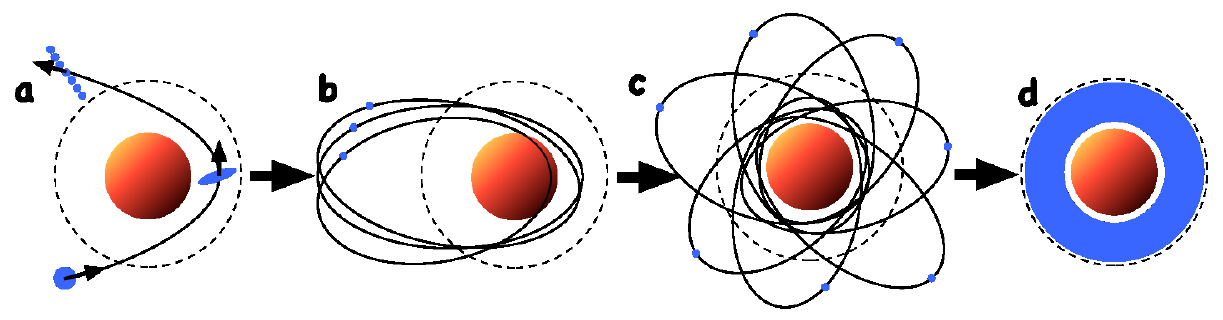}
 \vspace{0mm}
 \caption{Schematic illustration of our ring formation scenario via tidal disruption. Dashed line represents the Roche limit of the planet. (a) A large passing object (blue) experiences a close encounter with a giant planet (brown) and is tidally destroyed. (b) As a result of tidal disruption, some fragments are gravitationally captured into orbits with large eccentricities. (c) Planetary tides precess the orbits of captured fragments and form a torus-like structure, and thus orbital crossing occurs and highy energetic collisions are enhanced. (d) Due to collisional damping/grinding as well as successive tidal destructions, fragments settle into thin equatorial circular rings around the planet.}
 \label{summary}
\end{figure}

\begin{figure}
  \epsscale{1.00}
  \plotone{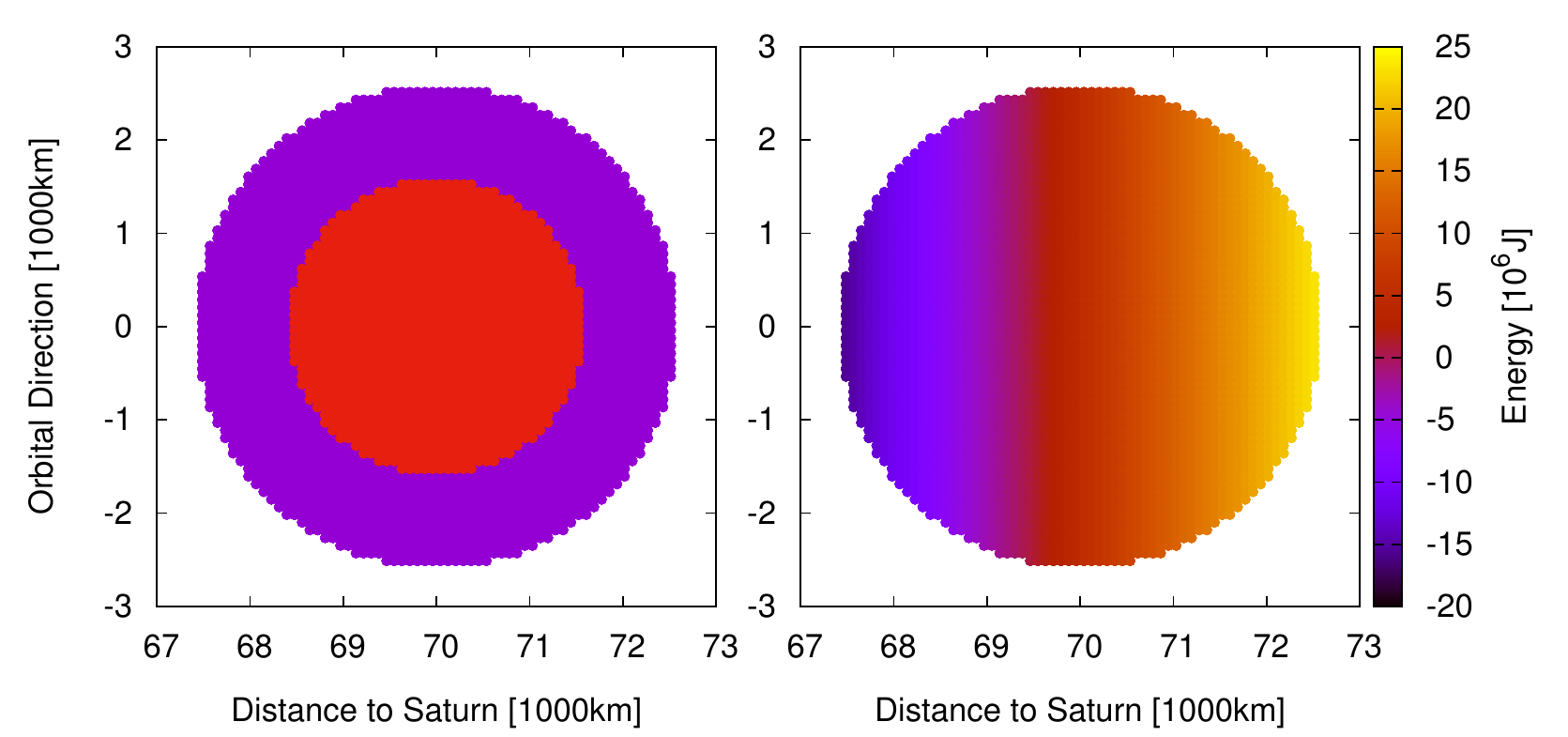}
  \caption{Internal structure (left panel) and the distribution of total energy (right panel) within a passing body on a hyperbolic orbit to Saturn. The body is at the pericenter at $7\times10^{4}$km seen from the normal direction to the orbital plane in the center of mass frame of the object coordinate and the relative velocity at infinity is $v_{\rm inf}=3$km/s. Saturn is to the left. On the left panel, different colours represent different densities of the components as red and blue, corresponding to rock and ice, respectively. On the right panel, the colour contour represents the total energy (kinematic energy + potential energy) of a cell.}
 \label{modeled_body}
\end{figure}

\begin{figure}
 \epsscale{1.0}
  \plotone{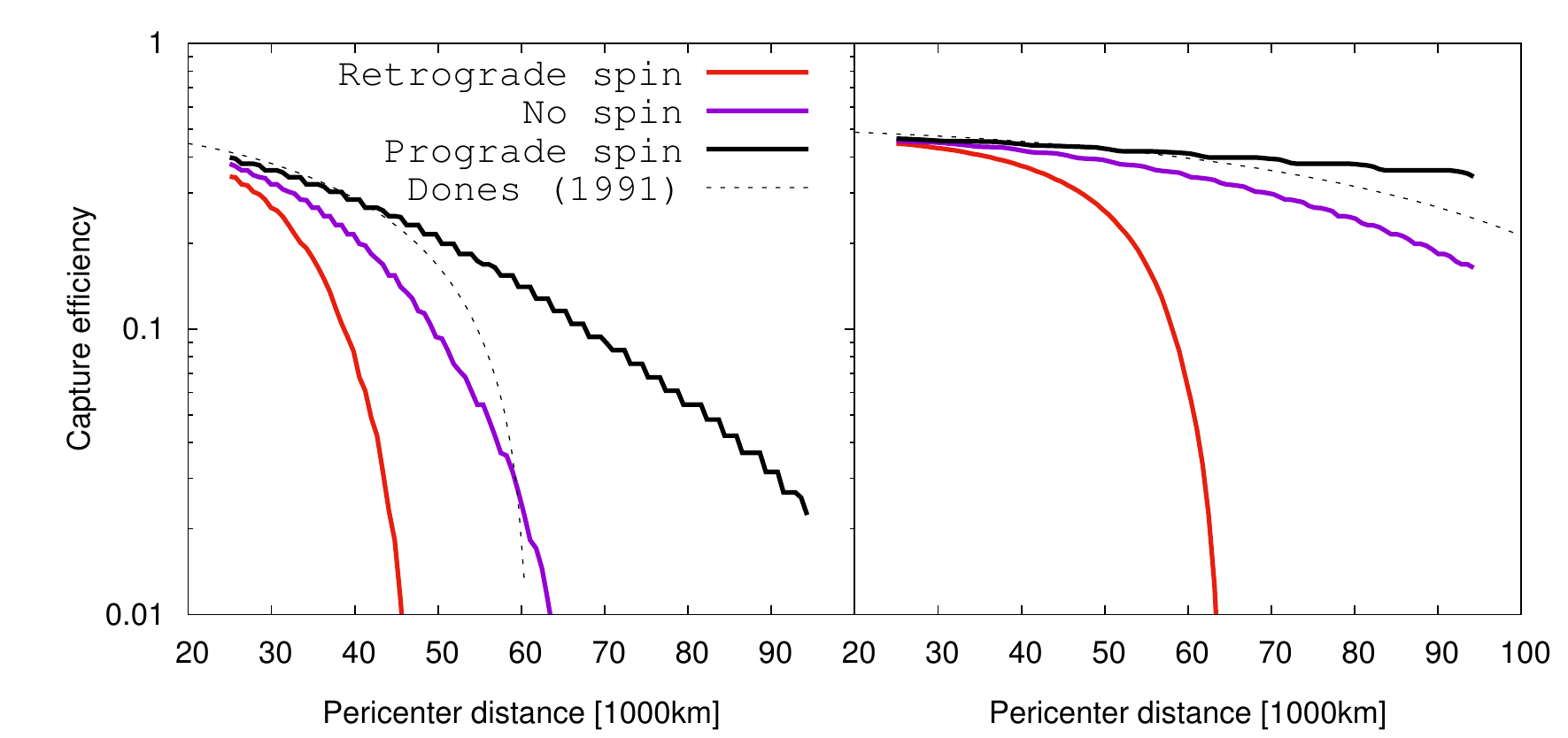}
 \caption{Capture efficiency ($M_{\rm cap}/M_{\rm body}$) at different pericenter distances, including different spin states of the body. Left and right panels show the case of $M_{\rm body}=10^{21}$kg and $10^{23}$kg, respectively. In the panels, black and red lines represent prograde and retrograde spins with the spin period $T_{\rm spin}=8$h, respectively. Purple line is the case of no spin.}
 \label{effect_of_spin}
\end{figure}

\begin{figure}
 \epsscale{1.0}
  \plotone{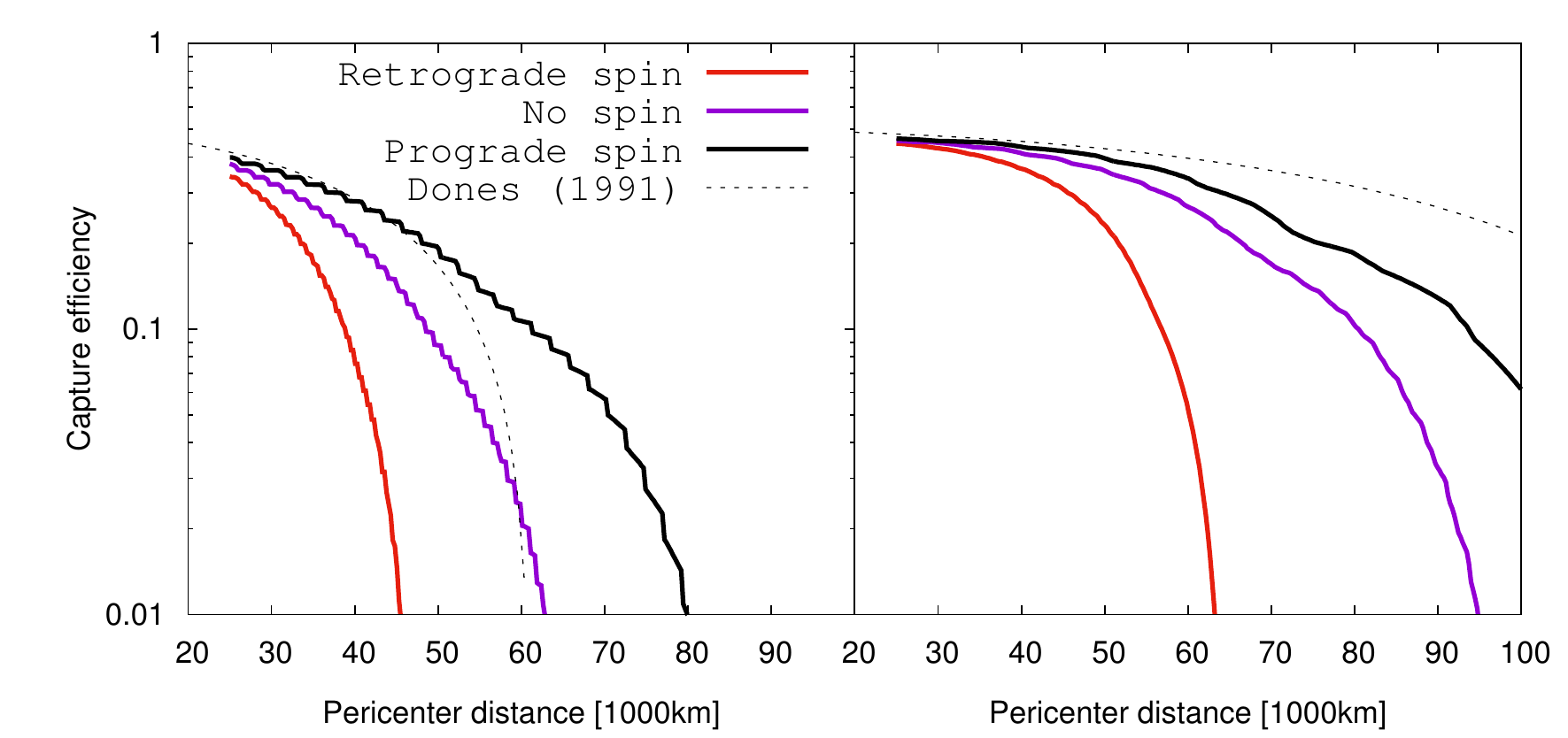}
 \caption{Same as Figure \ref{effect_of_spin} but includes the effects of both spin and self-gravity called "Hill capture".}
 \label{effect_of_grav}
\end{figure}

\begin{figure}
 \epsscale{1.0}
  \plotone{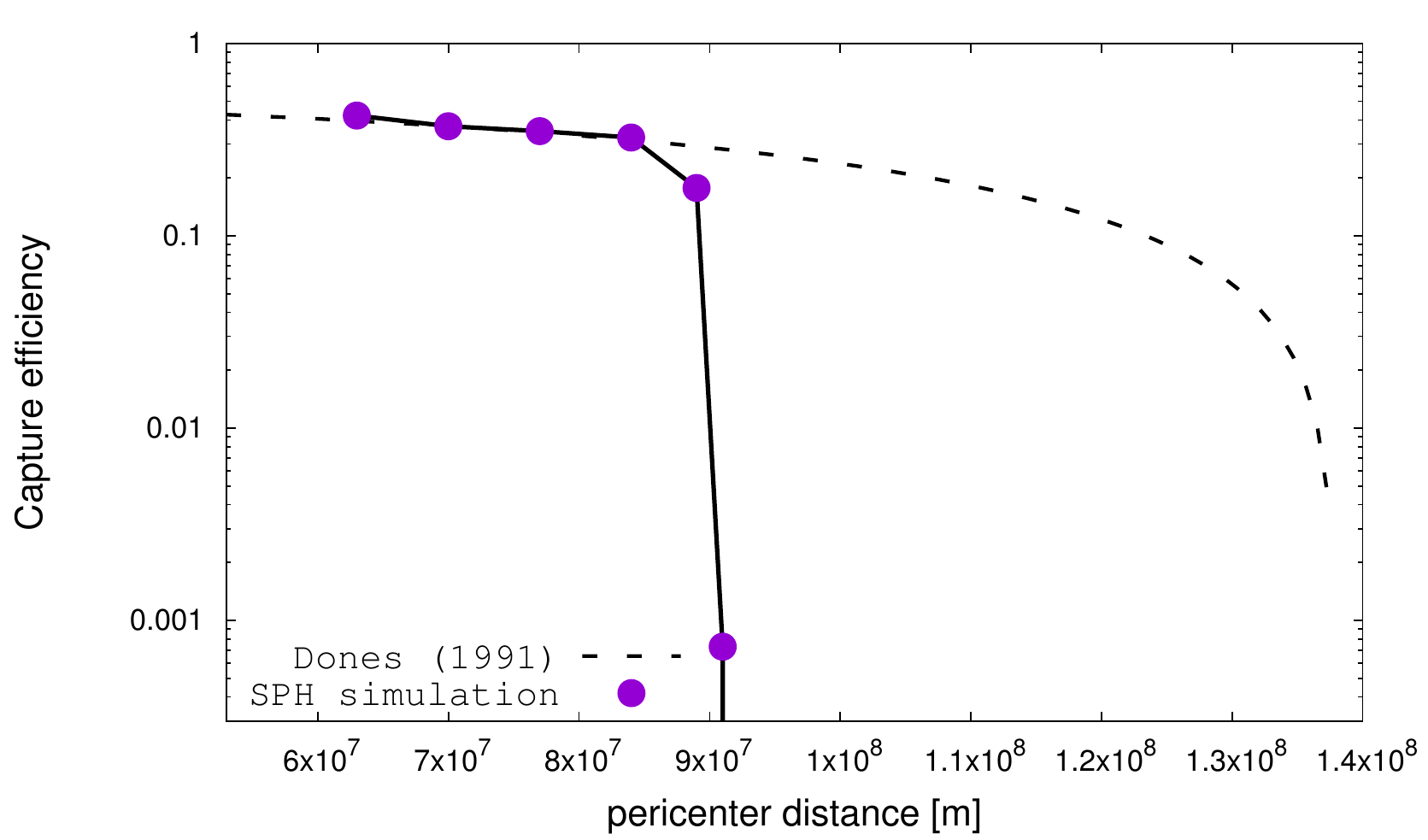}
 \caption{Capture efficiency ($M_{\rm cap}/M_{\rm body}$) of a homogeneous body ($M_{\rm body}=1 \times 10^{23}$kg) that experiences a close encounter with Saturn with $v_{\rm inf}=3$km/s.  Blue dots are obtained from SPH simulations and dashed line represents Dones' formula (Equation \ref{Dones1991}).}
 \label{homogeneous}
\end{figure}

\begin{figure}
 \epsscale{0.8}
  \plotone{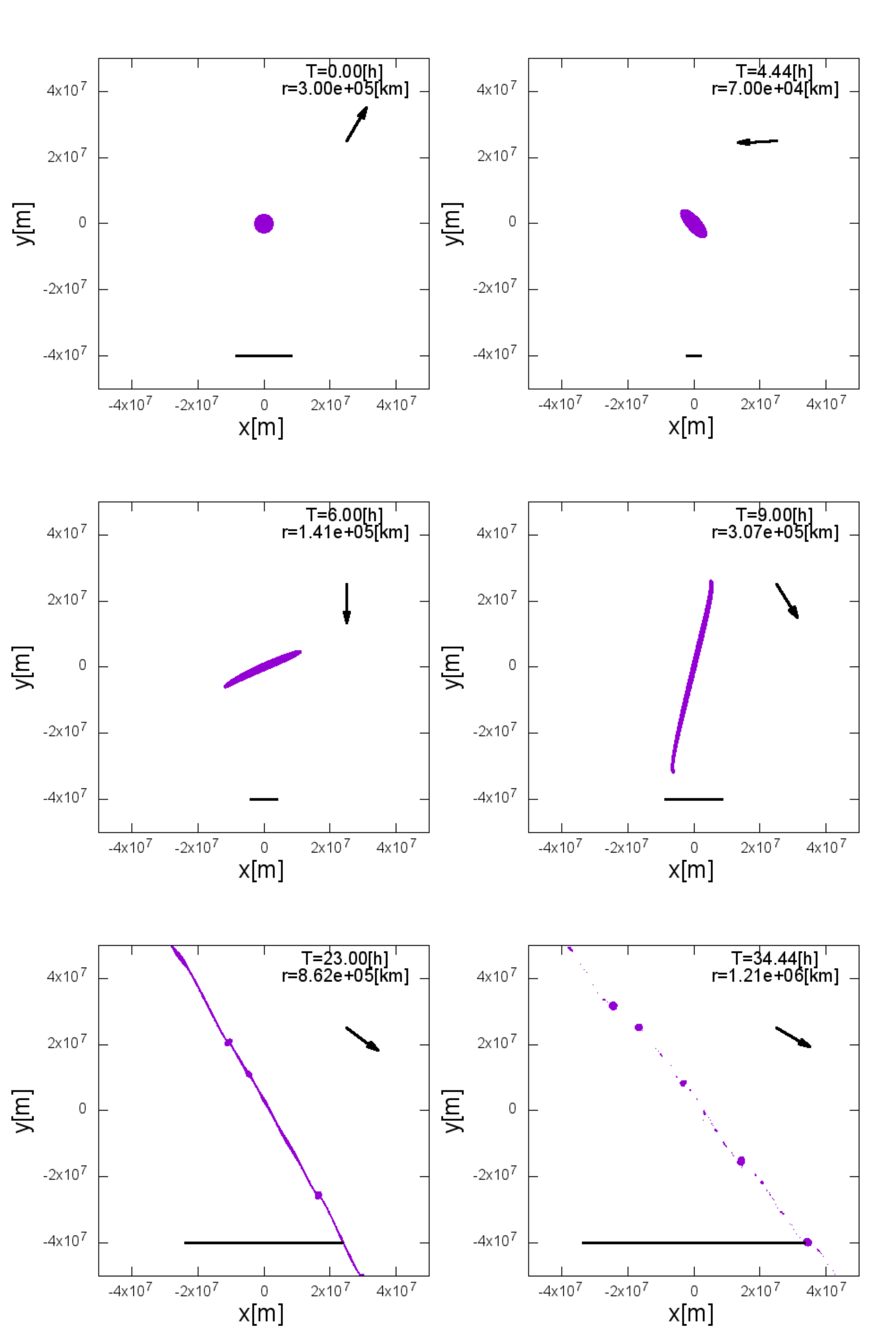}
 \caption{Snapshots of tidal disruption of $M_{\rm body}=1 \times 10^{23}$kg homogeneous body ($q=7.0 \times 10^{7}$m and $v_{\rm inf}=3$km/s) seen from the normal direction to the orbital plane in the center of mass frame of the object coordinate. The right top arrow points towards Saturn. The black horizontal line shows the Hill radius considering the object that has the initial mass. The distance to Saturn as well as time since the simulation started are also shown.}
 \label{homo_q7-0}
\end{figure}

\clearpage
\begin{figure}
 \epsscale{0.8}
  \plotone{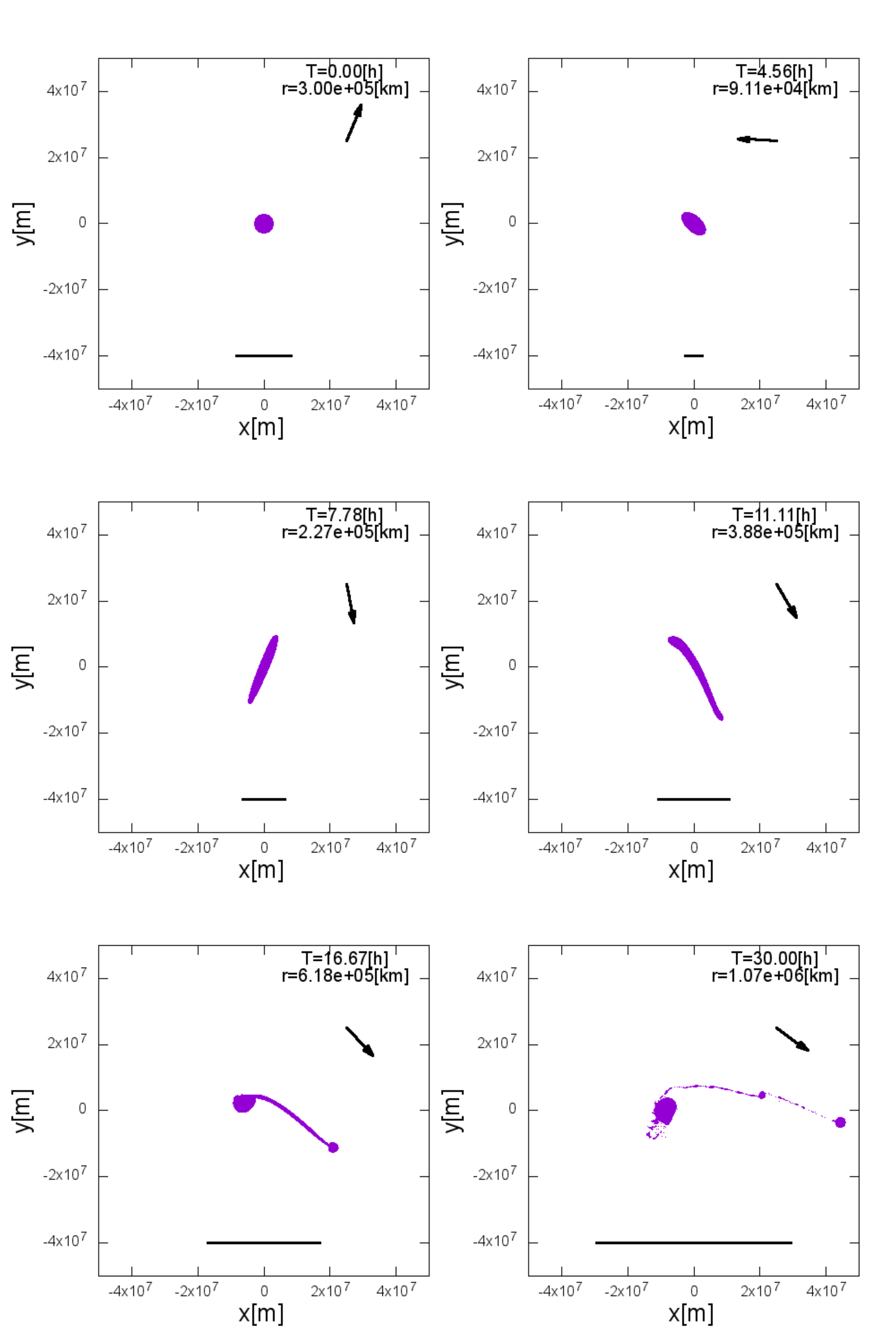}
 \caption{Same as Figure \ref{homo_q7-0} but case for $q=9.1 \times 10^{7}$m and $v_{\rm inf}=3$km/s.}
 \label{homo_q9-1}
\end{figure}

\begin{figure}
 \epsscale{1.0}
  \plotone{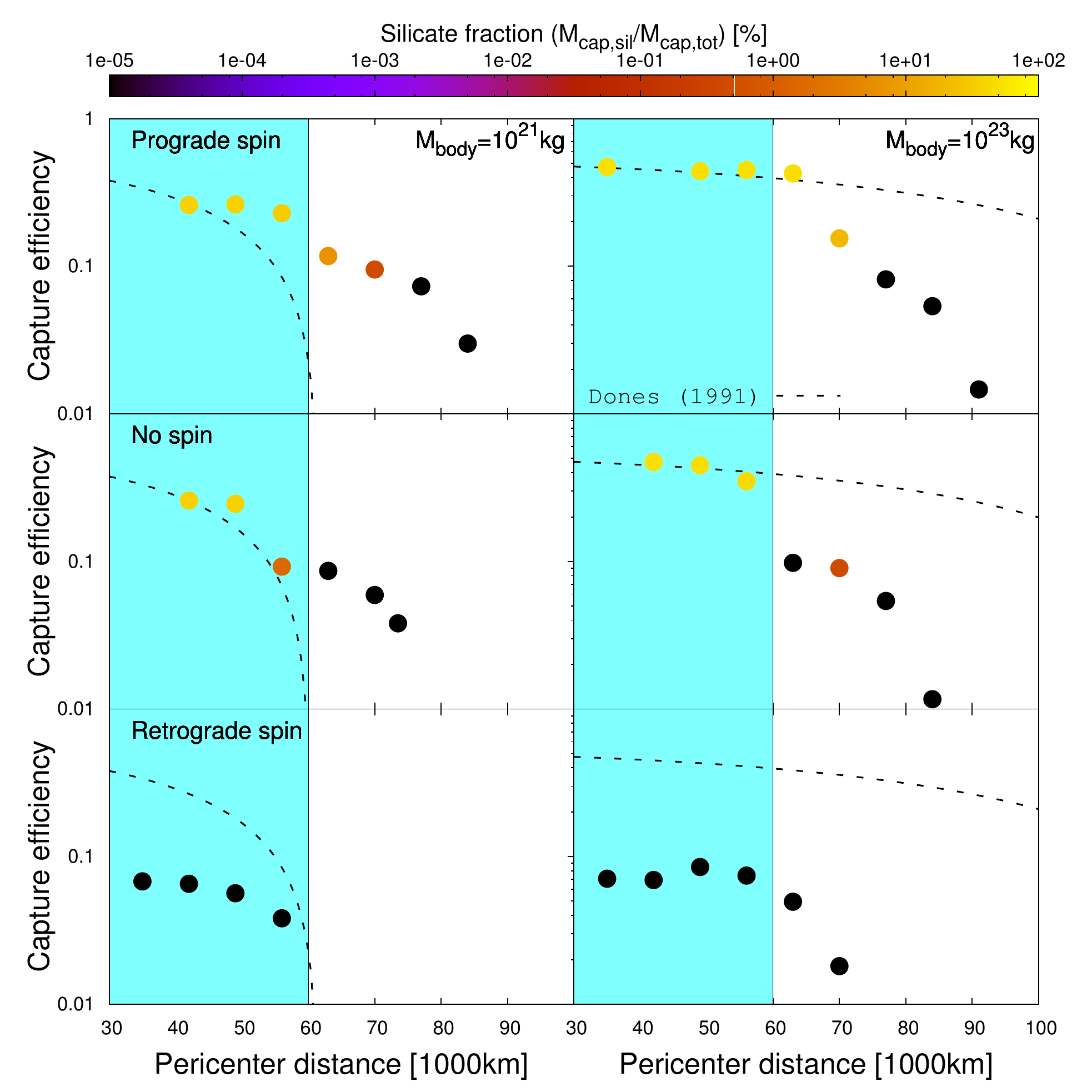}
 \caption{Capture efficiency ($M_{\rm cap,tot}/M_{\rm body}$) around Saturn. Data obtained from SPH simulations are shown by filled dots with the colour representing captured silicate fraction ($M_{\rm sil,cap}/M_{\rm cap,tot}$). Dones' formula is represented by dashed line. Left and right panels show cases of $M_{\rm body}=10^{21}, 10^{23}$kg bodies, respectively. From top to bottom panels, the case of initial prograde spin ($T_{\rm spin}=8$h), no spin, retrograde spin ($T_{\rm spin}=8$h) are shown, respectively. The light blue region corresponds to the region inside Saturn's radius.}
 \label{capture_efficiency_Saturn}
\end{figure}

\begin{figure}
    \epsscale{.8}
  \plotone{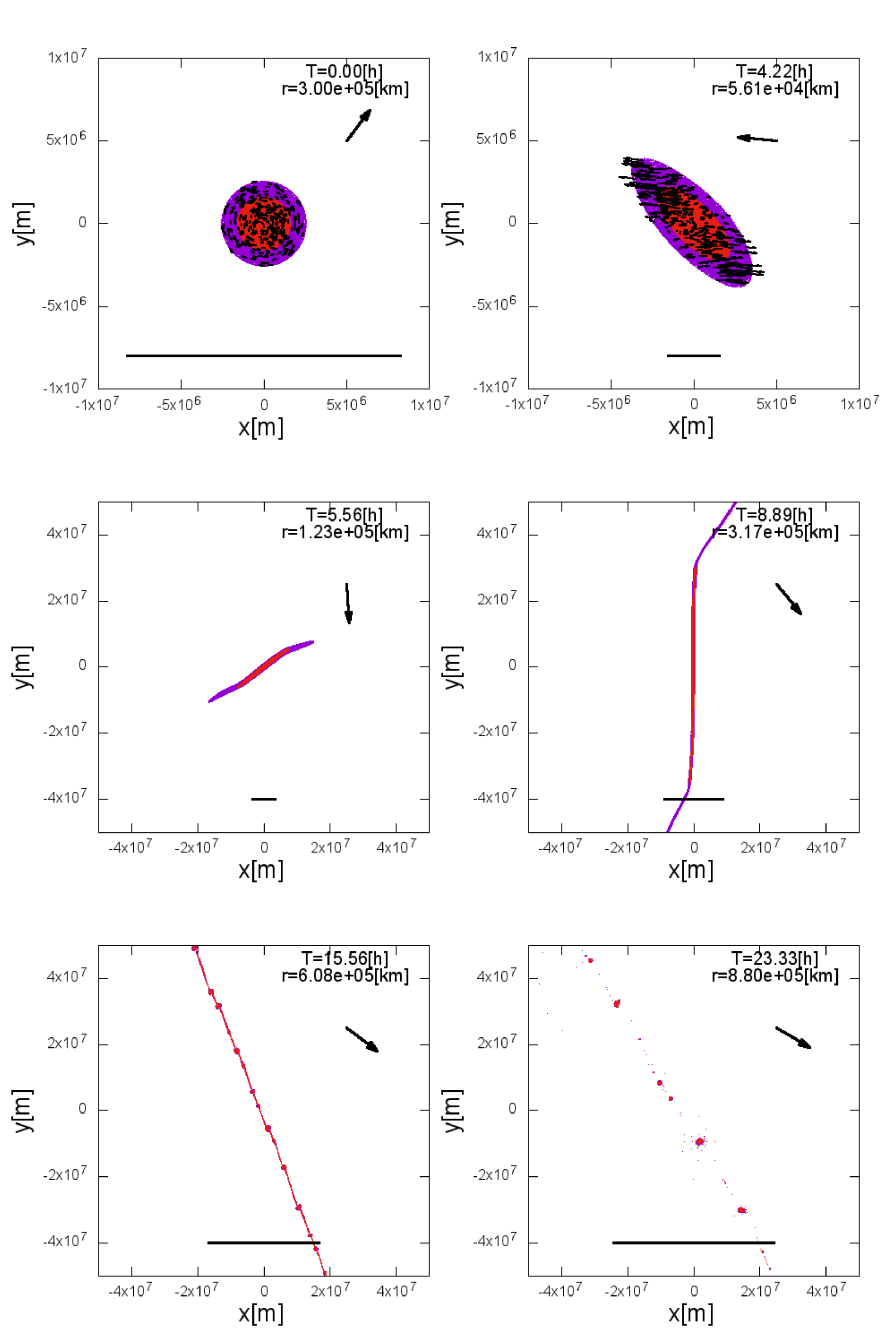}
 \caption{Snapshots of tidal disruption of a $M_{\rm body}=1 \times 10^{23}$kg differentiated body (initial prograde spin with the spin period $T_{\rm spin}=8$h, $q=5.6 \times 10^{7}$m and $v_{\rm inf}=3$km/s), seen from the normal direction to the orbital plane in the center of mass frame of the object coordinate. Red colour represents silicate core and blue colour represents icy mantle. The top right arrow points towards Saturn. Bottom black horizontal line is the Hill radius of object assuming initial mass. Distance to Saturn as well as time are also shown. On the top two panels, the velocity of particles in the center of mass frame are shown for some particles.}
 \label{differ_T-8h_q5-6}
\end{figure}

\begin{figure}
    \epsscale{.8}
  \plotone{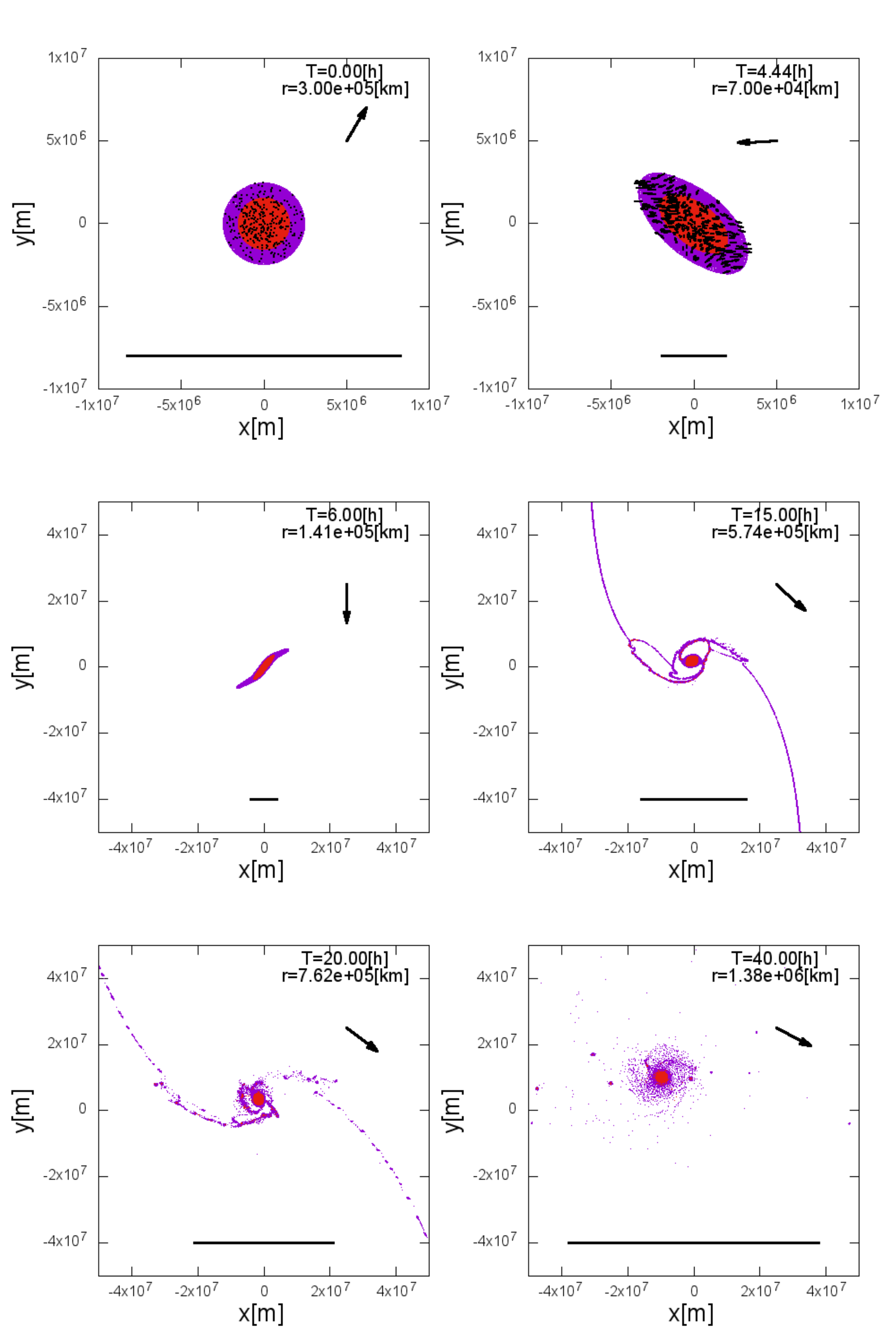}
 \caption{Same as Figure \ref{differ_T-8h_q5-6} but for a body with $M_{\rm body}=1 \times 10^{23}$kg, $T_{\rm spin}=\infty$, $q=7.0 \times 10^{7}$m and $v_{\rm inf}=3$km/s}
 \label{differ_Tinf_q7-0}
\end{figure}

\begin{figure}
    \epsscale{.8}
 \plotone{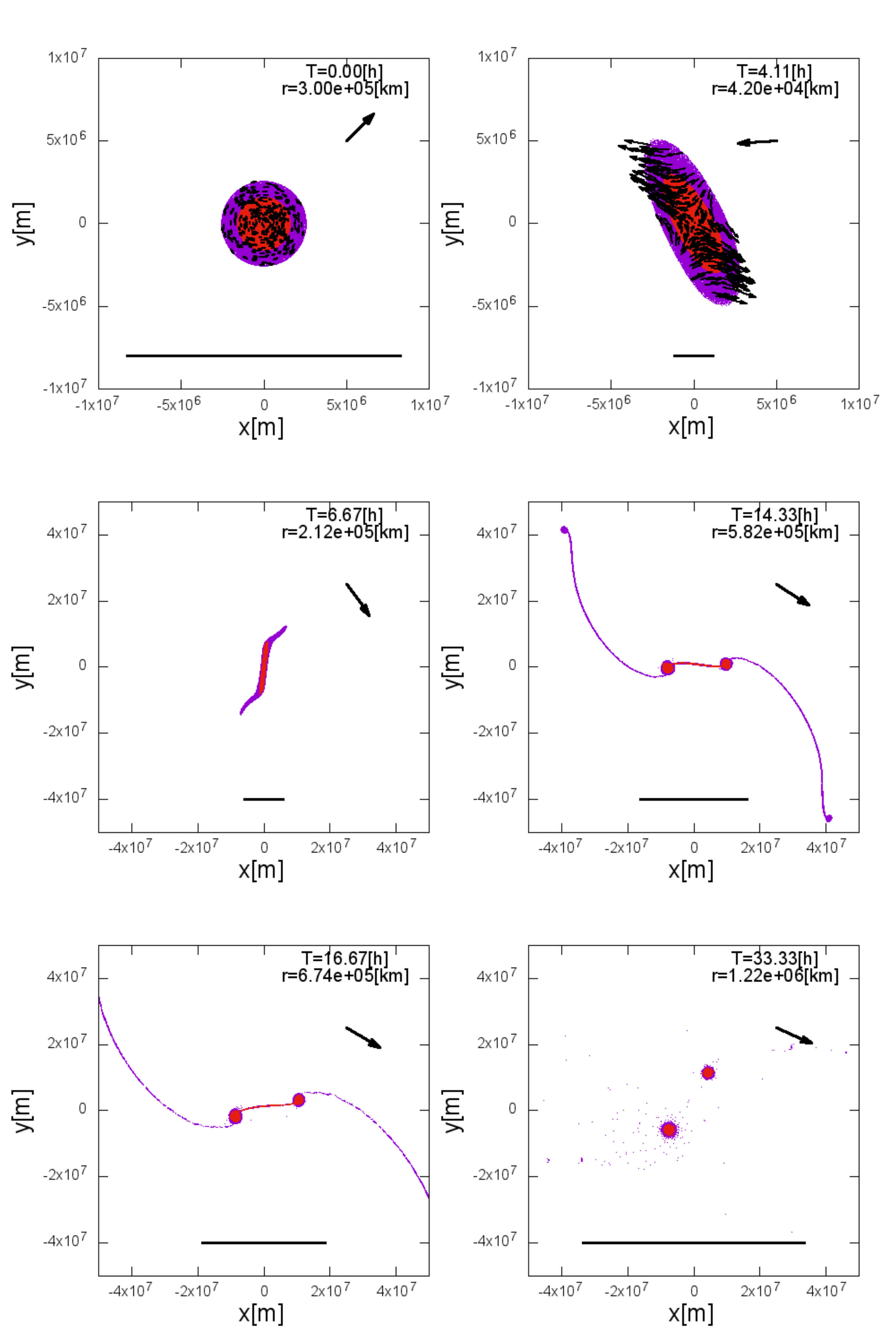}
 \caption{Same as Figure \ref{differ_T-8h_q5-6} but for a body with $M_{\rm body}=1 \times 10^{23}$kg, retrograde spin with $T_{\rm spin}=8$h, $q=4.2 \times 10^{7}$m and $v_{\rm inf}=3$km/s}
 \label{differ_T8h_q4-2}
\end{figure}

\begin{figure}
 \epsscale{1.0}
  \plotone{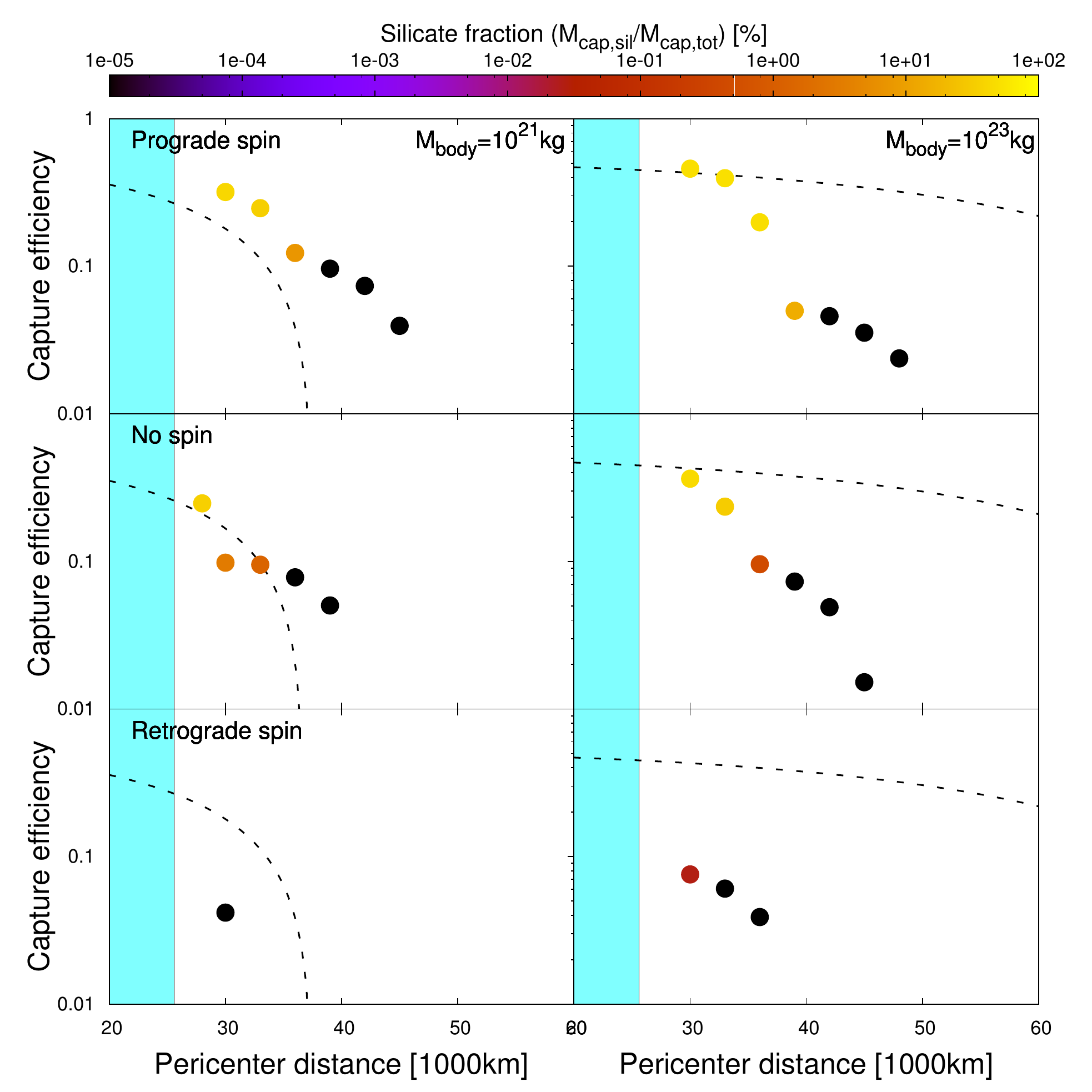}
 \caption{Same as Figure \ref{capture_efficiency_Saturn} but for Uranus.}
 \label{capture_efficiency_Uranus}
\end{figure}

\begin{figure}
\epsscale{1.0}
 \plotone{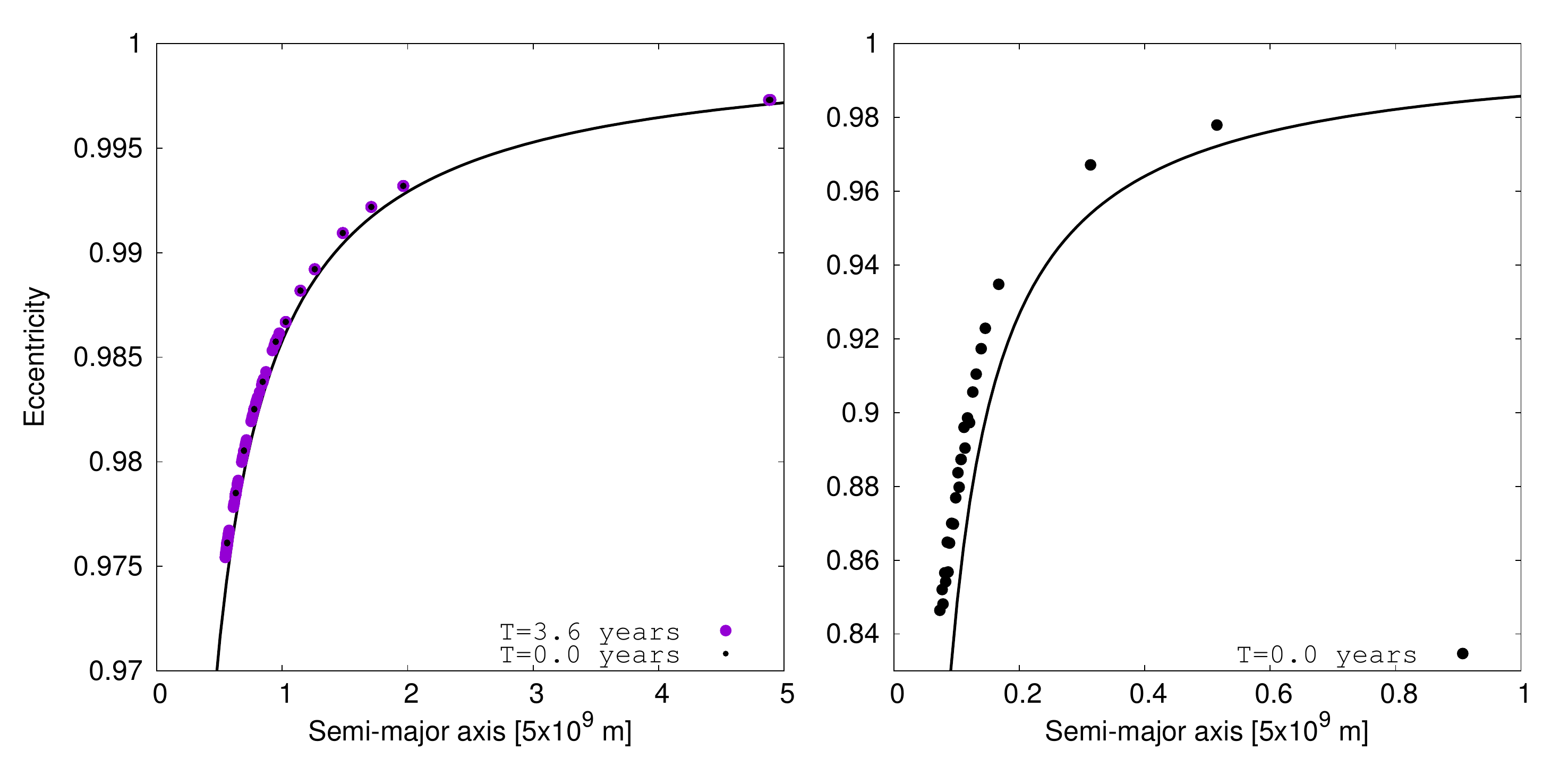}
 \caption{Eccentricities of captured fragments against their semi-major axis as well as an analytical estimation shown by a black curve (Equation \ref{ea_capture_analytic}) in the case of $M_{\rm body}=10^{21}$kg (left panel) and $10^{23}$kg (right panel), respectively (prograde spin with $T_{\rm spin}=8$h, $q=7.0 \times 10^{7}$m and $v_{\rm inf}=3.0$km/s around Saturn). Black dots are those obtained when SPH simulation is terminated ($T=0$ year for N-body simulation). Purple dots in the left panel are those obtained by N-body simulation at $T=3.6$ years.}
  \label{ea_fragments}
\end{figure}

\begin{figure}
 \epsscale{1.0}
  \plotone{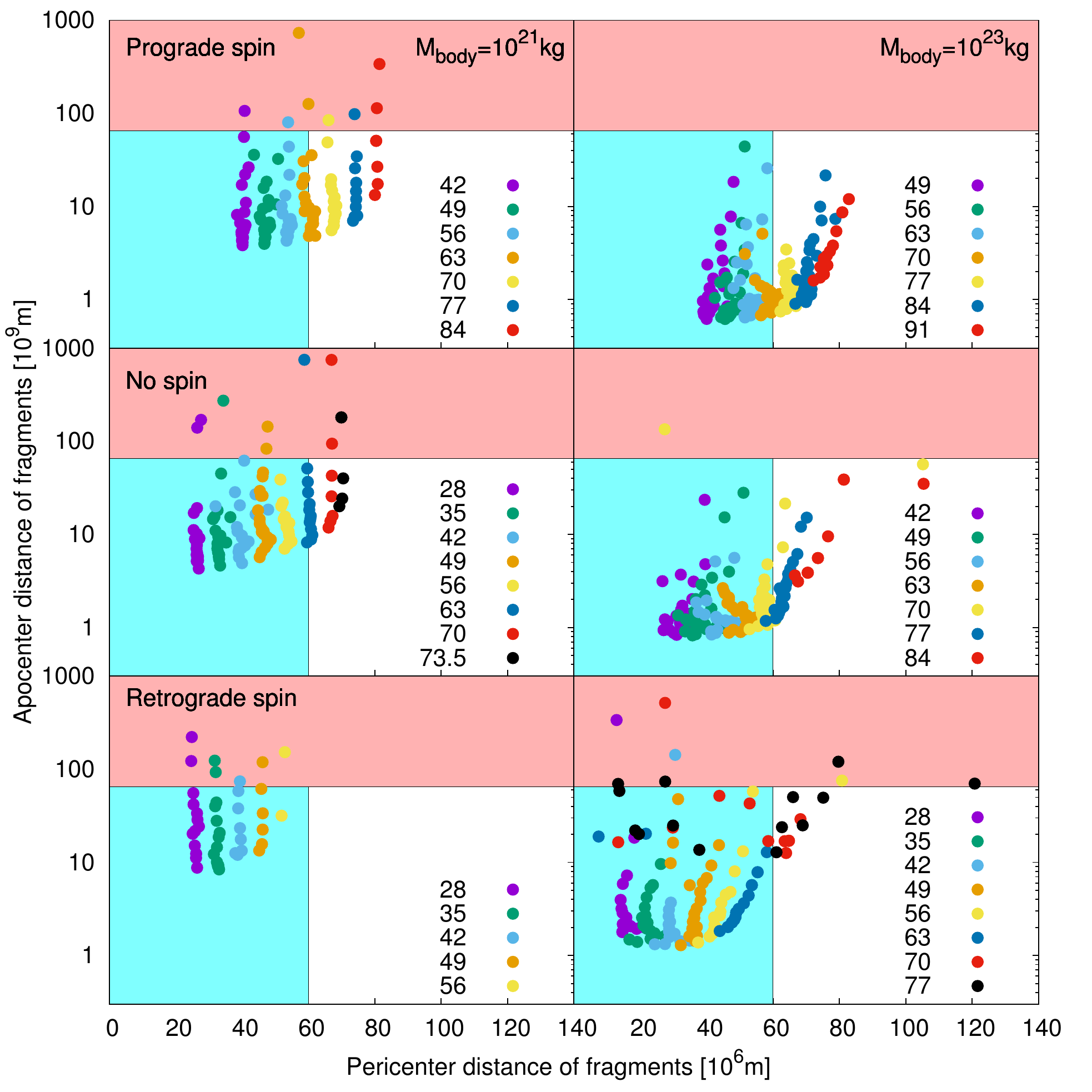}
 \caption{Pericenter and apocenter distances of the center of mass of the captured fragments around Saturn. Left and right panels show the results for $M_{\rm body}=10^{21}$kg and $10^{23}$kg, respectively. The light blue region is within Saturn's radius and light red region is outside the Hill radius of Saturn.  Different colours show the different results from a single SPH calculation with different initial pericenter distances of the passing object that is on a hyperbolic orbit (in units of $10^6$m). }
 \label{pericenter_apocenter_Saturn}
\end{figure}

\begin{figure}
 \epsscale{1.0}
 \plotone{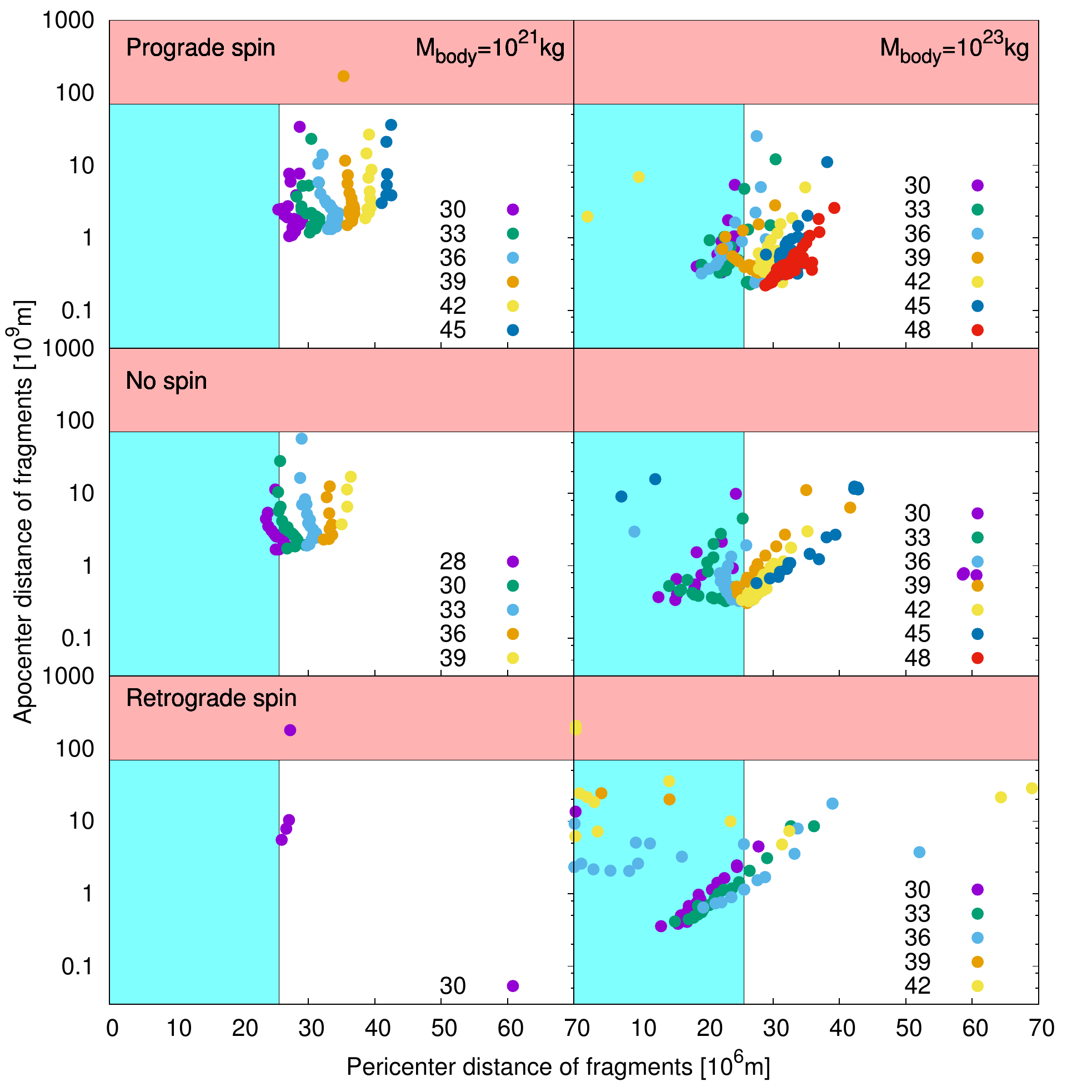}
 \caption{Same as Figure \ref{pericenter_apocenter_Saturn} but for Uranus.}
 \label{pericenter_apocenter_Uranus}
\end{figure}

\begin{figure}
 \epsscale{1.0}
  \plotone{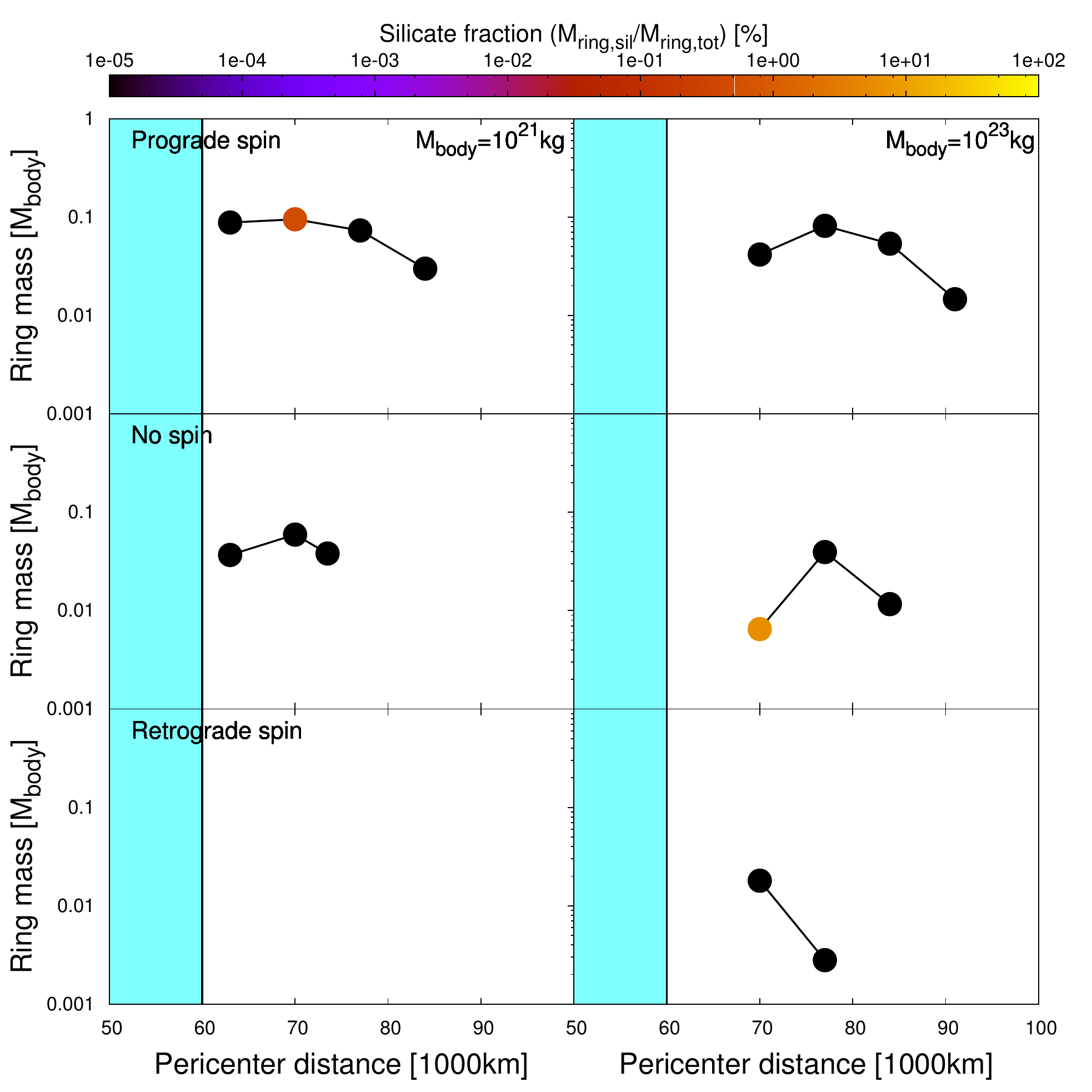}
 \caption{Estimated mass that can be incorporated into the rings around Saturn ($M_{\rm ring,tot}$). Each dot represents the data obtained from a single SPH calculation. Different colours show the silicate fraction in the estimated ring mass ($M_{\rm ring,sil}/M_{\rm ring,tot}$). The light blue region corresponds to the region inside Saturn's radius.}
 \label{Ring_mass_Saturn}
\end{figure}

\begin{figure}
 \epsscale{1.0}
  \plotone{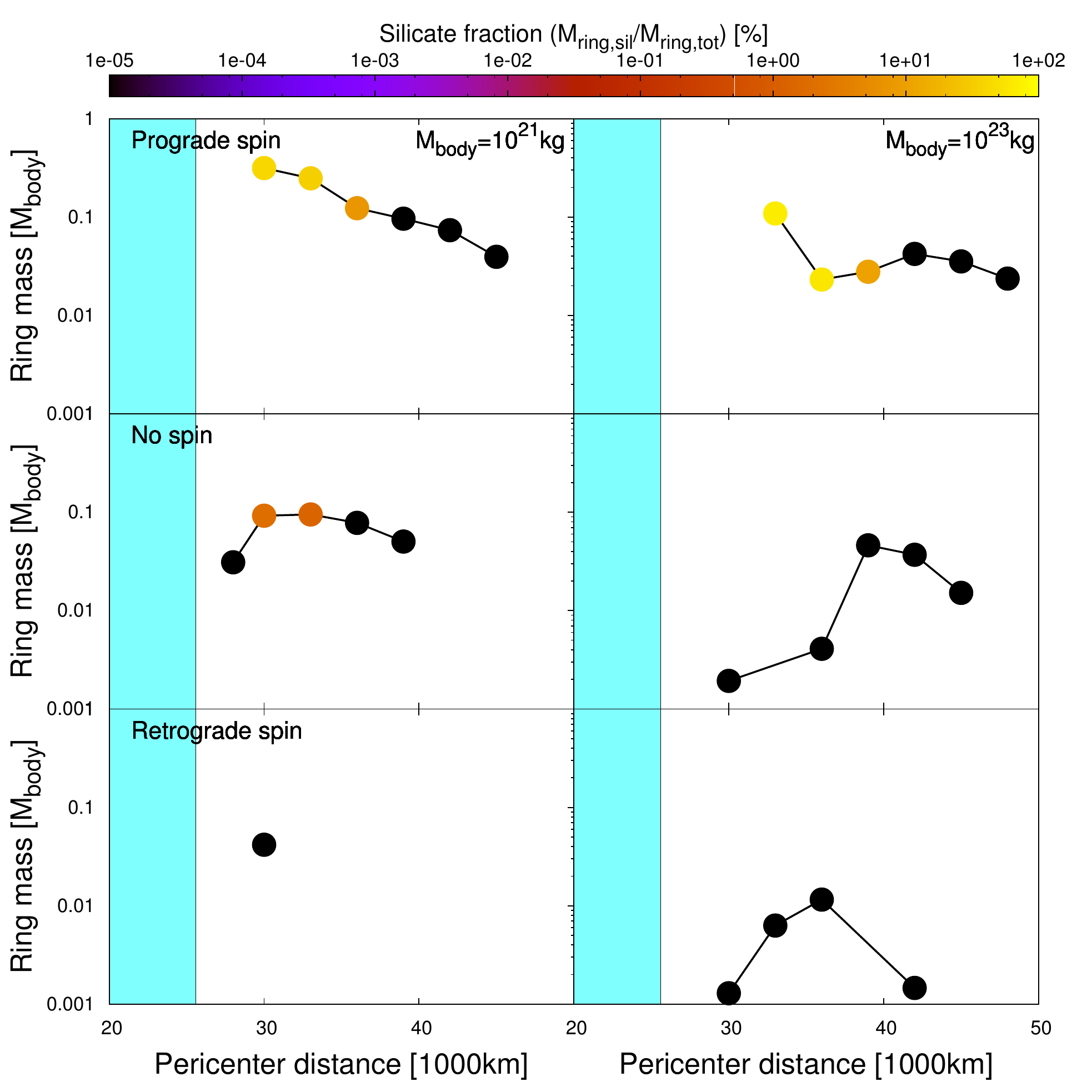}
 \caption{Same figure as Figure \ref{Ring_mass_Saturn} but in the case of Uranus.}
  \label{Ring_mass_Uranus}
\end{figure}

\begin{figure}
 \epsscale{1.0}
 \plotone{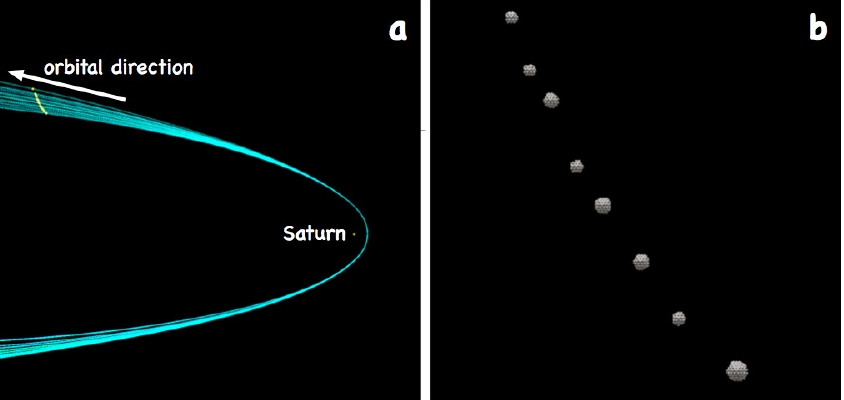}
 \caption{Initial state of our N-body simulations by using the data obtained from a single SPH simulation ($M_{\rm body}=1 \times 10^{21}$kg, prograde spin with $T_{\rm spin}=8$h, $q=7.0 \times 10^{7}$m and $v_{\rm inf}=3$km/s around Saturn). The total mass of the initial captured fragments is $M_{\rm cap}=9.5 \times 10^{19}$kg. In the left panel, yellow points show the initial positions of every particle and Saturn, and blue curves correspond to the particles' eccentric orbits. The right panel shows the modelled fragments on their initial orbits.}
  \label{Nbody_initial}
\end{figure}

\begin{figure}
\epsscale{1.0}
 \plotone{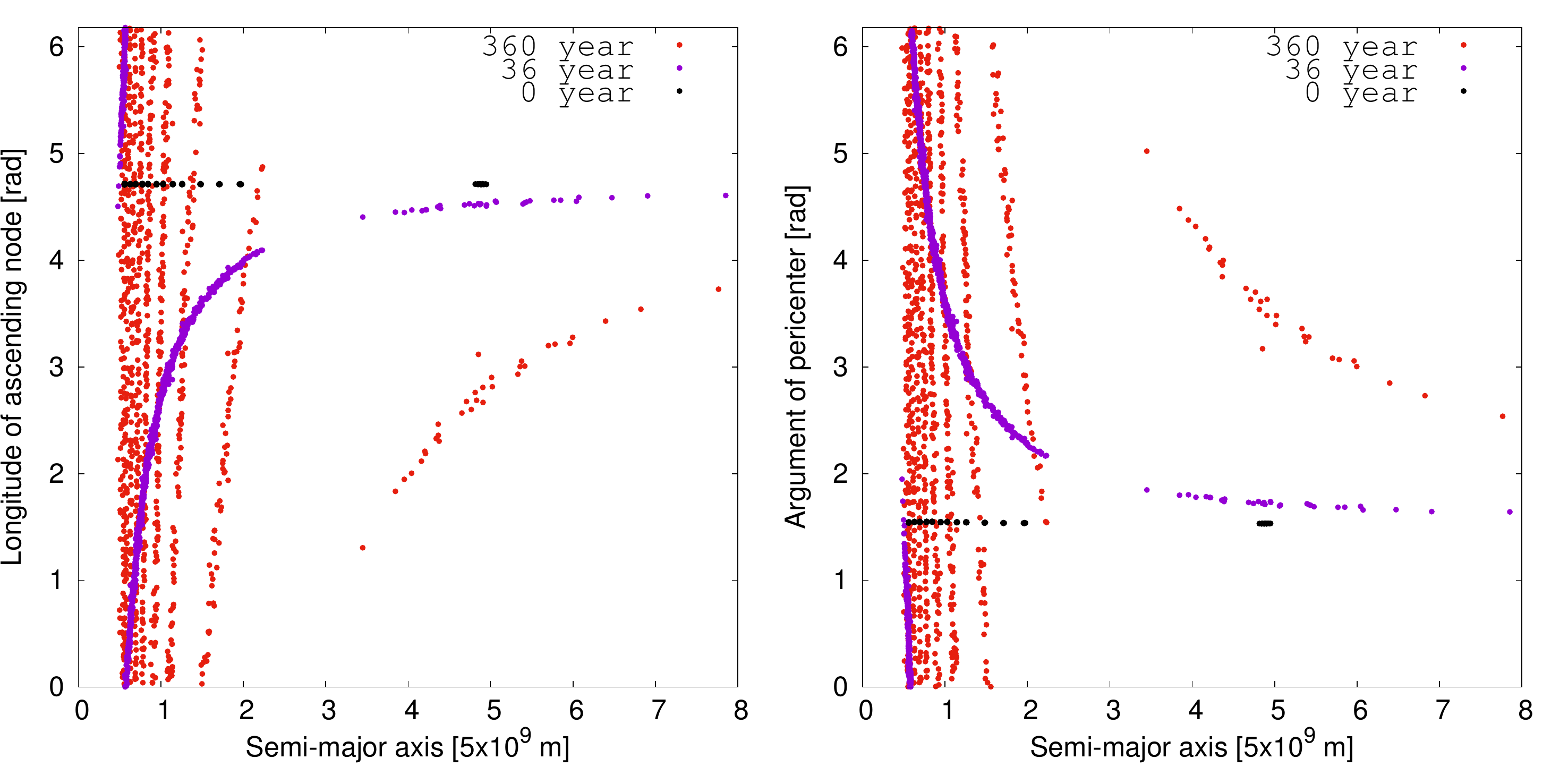}
 \caption{Left and right panels show the particles' longitudes of ascending node and arguments of pericenter, respectively. Particles with smaller semi-major axis are randomised more quickly. Note that in this calculation collisions are not considered.}
  \label{Nbody_no_collision}
\end{figure}

\begin{figure}
 \epsscale{1.0}
 \plotone{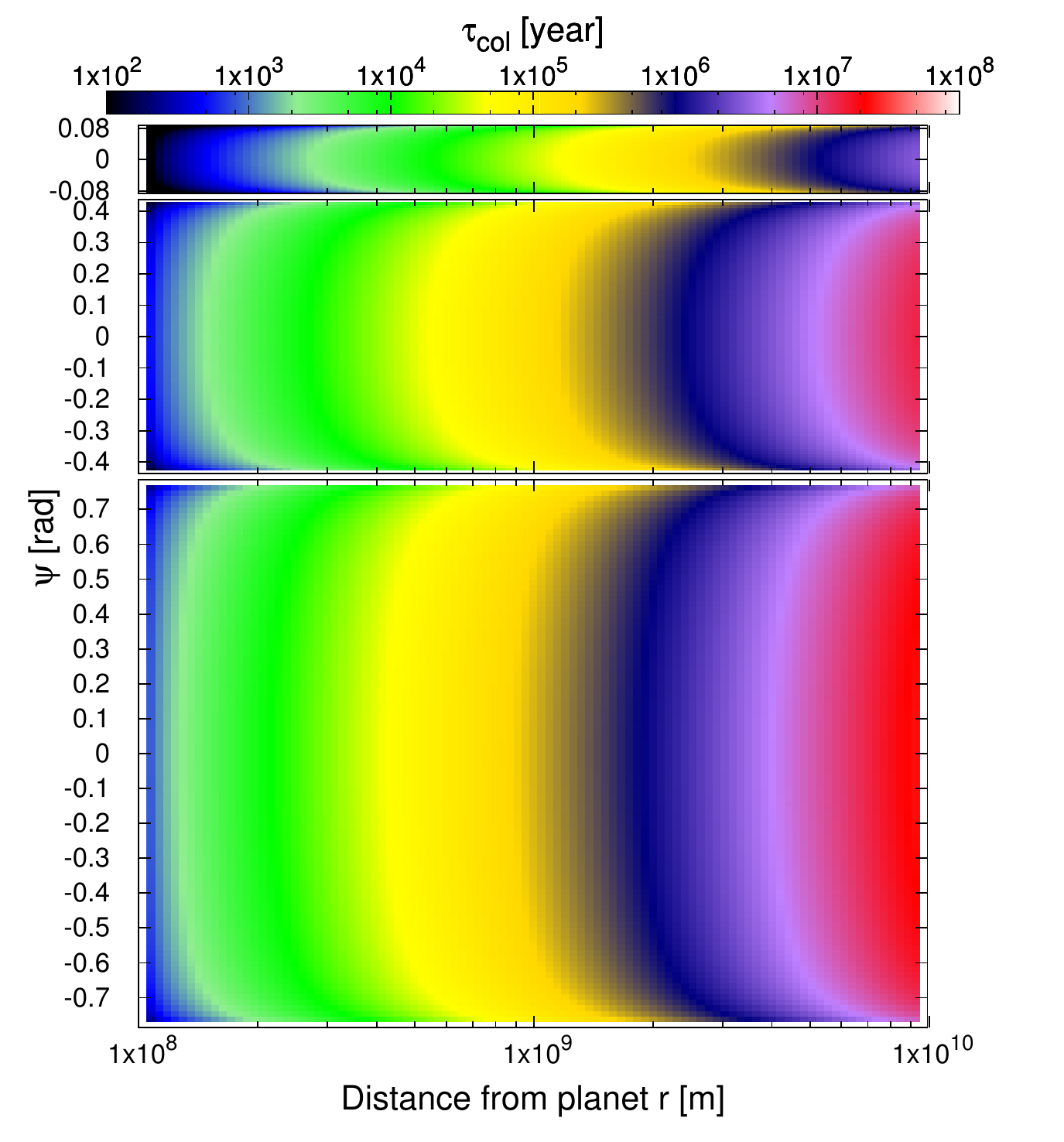}
 \caption{Collision timescale estimated by Equation (\ref{timescale}) with $a=5.0\times10^{9}$m, $e=0.98$, $i=5$ (top panel), $25$ (middle panel) and $45$ (bottom panel) degrees, $r_{\rm p}=26$km and $N_{\rm tot}$=1000. The colour indicates the collisional timescale in units of years.}
  \label{tau_coll_Seb}
\end{figure}

\begin{figure}
 \epsscale{1.0}
 \plotone{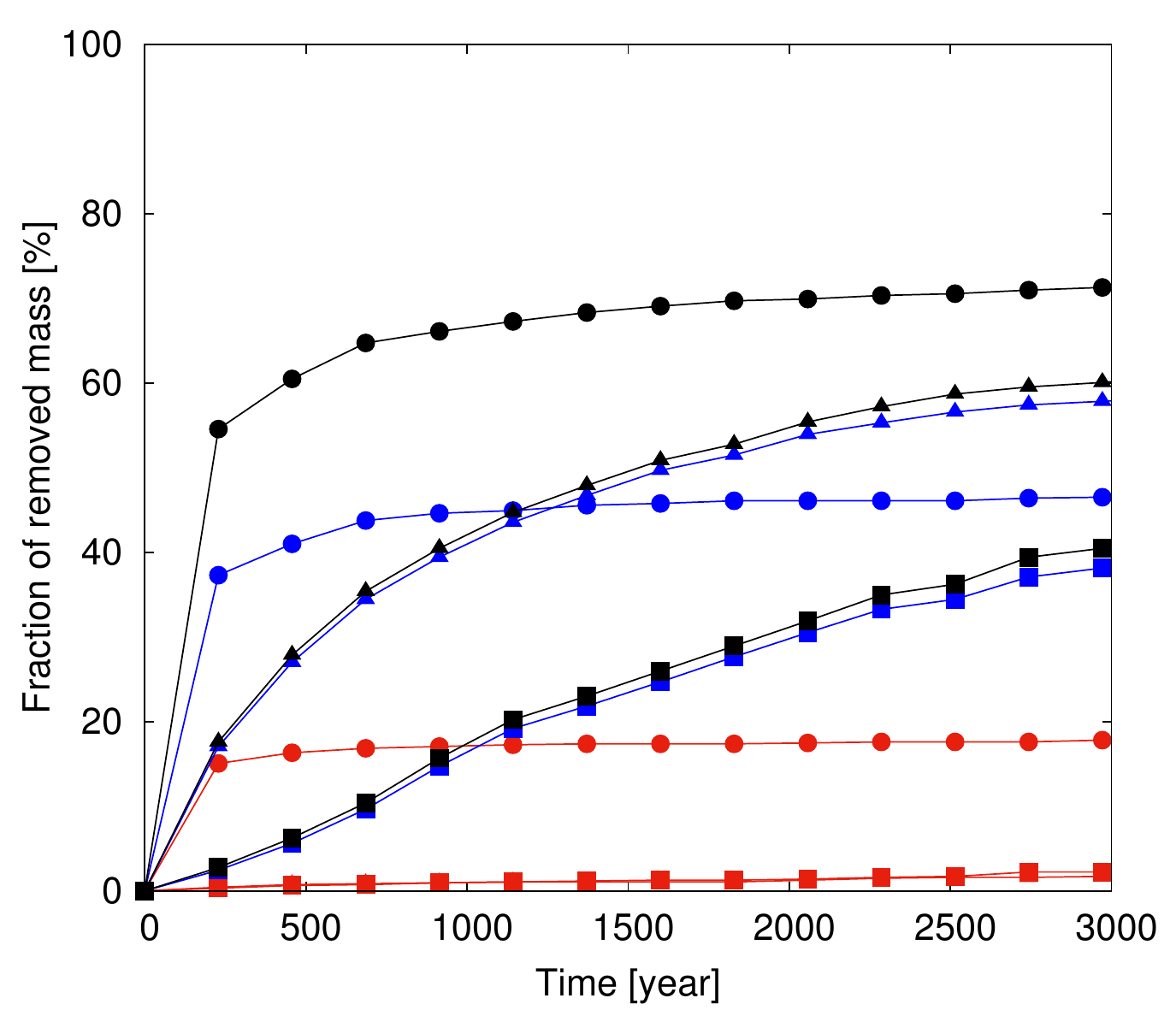}
 \caption{Fraction of removed mass due to the effect of Titan. Different symbols represent different initial orbital inclinations of captured fragments with respect to Titan's orbital plane; circles, triangles and squares represent inclination of $0$, $25$ and $45$ degrees, respectively. The black marks with lines are the total fraction of removed mass (those accreted by either Saturn or Titan, or those go beyond 10 times of Saturn's Hill radius). The blue and red marks with lines are fractions of captured mass that is accreted by Saturn and Titan, respectively.}
  \label{effect_of_titan}
\end{figure}

\end{document}